\documentclass[12pt,english,floatfix,nofootinbib,superscriptaddress,aps,prd,preprint]{revtex4-2}
\usepackage[utf8]{inputenc}       
\usepackage[english]{babel}       
\usepackage[utf8]{inputenc}
\usepackage{textcomp}
\usepackage{amsmath,amsthm,amssymb,amsfonts,mathrsfs,amsbsy} 
\usepackage{tensor}               
\usepackage{slashed}  
\usepackage{bbm}
\usepackage{esint}    
\usepackage[a4paper, margin=1.6cm]{geometry}
\usepackage{cancel}    
\usepackage{caption}
\usepackage{graphicx}             
\usepackage{natbib}
\usepackage{float}                
\usepackage{subfig}               
\usepackage[font=small,labelfont=bf]{caption} 
\usepackage{multirow}             
\usepackage{array}                
\usepackage{tikz}                 
\usetikzlibrary{quotes,angles,arrows,decorations.markings} 
\usepackage{makecell} 

\usepackage{lipsum}

\usepackage{xcolor}               
\usepackage{color}                
\usepackage{textcomp}             
\usepackage{units}                

\newcommand{\be}{\begin{equation}}
\newcommand{\ee}{\end{equation}}
\newcommand{\Be}{\begin{eqnarray}}
\newcommand{\Ee}{\end{eqnarray}}

\newcommand{\mincir}{\raise
-3.truept\hbox{\rlap{\hbox{$\sim$}}\raise4.truept\hbox{$<$}\ }}
\newcommand{\magcir}{\raise
-3.truept\hbox{\rlap{\hbox{$\sim$}}\raise4.truept\hbox{$>$}\ }}

\newcolumntype{Y}{>{\centering\arraybackslash}X}
\providecommand{\U}[1]

\usepackage[dvips]{epsfig}        
\usepackage[dvips]{graphicx}      
\usepackage{hyperref}             
\hypersetup{
    colorlinks=true,              
    breaklinks=true,              
    citecolor=blue,               
    linkcolor=[rgb]{0,0.5,0.9},   
    urlcolor=red,                 
    filecolor=green               
}

\newcommand{\ie}{\begin{equation}}
\newcommand{\fe}{\end{equation}}
\newcommand{\se}{\begin{eqnarray}}
\newcommand{\ff}{\end{eqnarray}}

\begin{document}

\title{A rotating black hole in a Hernquist dark matter halo: horizon geometry, thermodynamics, and quantum emission}


\author{A. A. Ara\'{u}jo Filho}
\email{dilto@fisica.ufc.br}
\affiliation{Departamento de Física, Universidade Federal da Paraíba, Caixa Postal 5008, 58051--970, João Pessoa, Paraíba,  Brazil.}
\affiliation{Departamento de Física, Universidade Federal de Campina Grande Caixa Postal 10071, 58429-900 Campina Grande, Paraíba, Brazil.}
\affiliation{Center for Theoretical Physics, Khazar University, 41 Mehseti Street, Baku, AZ-1096, Azerbaijan.}

\author{Arun Kumar}
\email{arunbidhan@gmail.com}
\affiliation{Centre for Theoretical Physics, Jamia Millia Islamia, New Delhi 110025, India.}

\author{N. Heidari}
\email{heidari.n@gmail.com}
\affiliation{Center for Theoretical Physics, Khazar University, 41 Mehseti Street, Baku, AZ-1096, Azerbaijan.}
\affiliation{School of Physics, Damghan University, Damghan, 3671641167, Iran.}
\affiliation{Departamento de Física, Universidade Federal de Campina Grande Caixa Postal 10071, 58429-900 Campina Grande, Paraíba, Brazil.}


\author{C. F. S. Pereira}
	\email{carlosfisica32@gmail.com}
	\affiliation{Departamento de F\'isica e Qu\'imica, Universidade Federal do Esp\'irito Santo, Av.Fernando Ferrari, 514, Goiabeiras, Vit\'oria, ES 29060-900, Brazil.}

\author{Amilcar R. Queiroz}
\email{amilcarq@df.ufcg.edu.br}

\affiliation{Departamento de Física, Universidade Federal de Campina Grande Caixa Postal 10071, 58429-900 Campina Grande, Paraíba, Brazil.}


\author{V. B. Bezerra}
\email{valdir@fisica.ufpb.br}
\affiliation{Departamento de Física, Universidade Federal da Paraíba, Caixa Postal 5008, 58051--970, João Pessoa, Paraíba,  Brazil.}

\date{\today}

\begin{abstract}

We investigate the geometrical, thermodynamic, and quantum emission properties of a rotating black hole immersed in a Hernquist dark matter halo. Starting from a static black hole spacetime surrounded by a Hernquist distribution, we construct its rotating counterpart through the noncomplexification formulation of the Newman--Janis algorithm and analyze the modifications induced by the {independent halo parameters $\rho$ and $r_s$} and the rotation parameter $a$. The horizon structure is determined from the roots of the radial function $\Delta(r)$, while the stationary limit surfaces and the corresponding ergoregions are obtained from the condition $g_{tt}=0$. We show that the Hernquist contribution displaces the outer event horizon toward larger radii and modifies the size of the ergoregion, whereas rotation controls the oblateness of the horizon and the strength of frame dragging. We further derive the surface gravity, Hawking temperature, Bekenstein--Hawking entropy, and heat capacity. The quantum tunneling rate is obtained from the Hamilton--Jacobi method, leading to the corresponding occupation number and {a thermal estimate of the particle creation density}. {Finally, we estimate the Hawking luminosity and evaporation timescales within a Stefan--Boltzmann approximation.} All standard Kerr and Schwarzschild results are recovered in the appropriate limiting cases.

\end{abstract}


\maketitle

\clearpage

\tableofcontents


\section{Introduction}

Black holes are among the most important predictions of General Relativity (GR), since they describe the strong--field regime in which gravity becomes fully nonlinear. In the classical picture, they may arise from the gravitational collapse of sufficiently massive compact objects, with the singularity theorems showing that spacetime singularities are a generic outcome under suitable energy and causality conditions \cite{Schwarzschild:1916uq,Penrose:1964wq,Wald:1984rg,dInverno:1992gxs}. The first exact solution of this kind was obtained by Schwarzschild and describes the exterior vacuum geometry of a static and spherically symmetric object. This result was later extended in different directions, including electrically charged configurations, cosmological backgrounds, and rotating geometries. In particular, the Kerr solution provides the reference model for an axisymmetric vacuum black hole, while the Kerr--Newman metric incorporates both rotation and electric charge \cite{Kerr:1963ud,Newman:1965my}. These geometries have become fundamental in the description of compact astrophysical objects and in the analysis of strong gravitational fields.

{Related investigations of nonstandard compact geometries have increasingly combined thermodynamic, dynamical, and optical diagnostics. For regular black holes, particle motion, shadows, and thermodynamic stability have been studied both in pure-gravity models and in geometries satisfying the limiting curvature condition \cite{Ditta:2024PureGravity,Ditta:2025LCC}. In matter-dressed configurations, thermal fluctuations, particle collisions, quasi-periodic oscillations, and emission rates have been examined for Schwarzschild black holes immersed in Dehnen-type dark matter halos \cite{Ashraf:2025Dehnen}, while the orbital dynamics and QPO properties of black holes surrounded by pseudo-isothermal dark matter halos have also been investigated \cite{Mustafa:2025PseudoIsothermal}. Entropy deformations provide another related direction. In particular, Barrow entropy has been analyzed together with thermodynamic stability, particle motion, emission rates, and QPOs for Frolov black holes, both in the absence and in the presence of a surrounding quintessence field \cite{Bouzenada:2025FrolovBarrow,Maurya:2025Barrow}. Complementary studies of wormhole geometries have developed geometric and wave-optical descriptions based on ray trajectories, the Helmholtz equation, effective potentials, and refractive indices. These include minimal-surface traversable wormholes, topologically charged Perry--Mann type wormholes with disclinations, and Lorentz-violating wormholes containing topological defects \cite{Dogan:2025MinimalSurface,Ahmed:2025PerryMann,Ahmed:2025LorentzWormhole}.}

The classical description of black holes changed substantially after the development of black hole thermodynamics. The area theorem, the analogy between surface gravity and temperature, and the identification of the horizon area with entropy showed that black holes obey laws closely related to the ordinary laws of thermodynamics \cite{Bekenstein:1973ur,Bardeen:1973gs,hawking1976black,Filho:2024ilq,filho2024implications,Kumar:2023gjt}. This connection became physical after Hawking demonstrated that black holes emit radiation due to quantum effects in the vicinity of the event horizon \cite{Hawking:1975vcx}. As a consequence, the horizon temperature is given by $T=\kappa/(2\pi)$, whereas the Bekenstein--Hawking entropy is proportional to the horizon area. These results established a precise link among GR, Quantum Field Theory in curved spacetime and Thermodynamics \cite{Gibbons:1977mu,Davies:1977bgr,Toussaint:1978br,AraujoFilho:2025rwr,Hawking:1982dh,York:1986it,Gibbons:1996af,Sood:2024ufi,Kumar:2024qon,Kumar:2024bls,AraujoFilho:2025hkm,Birrell:1982ix,Frolov:1998wf,Kumar:2025kzt,araujo2023thermodynamics,Page:2004xp,Carlip:2014pma,araujo2022thermal,Heidari:2025iiv}.

Rotating black holes are especially relevant from an astrophysical point of view, since angular momentum is expected to be present in realistic compact systems. Nevertheless, obtaining rotating geometries from nonvacuum static seeds is not a trivial task. A standard procedure is the Newman--Janis algorithm, originally formulated in terms of null tetrads and complex coordinate transformations \cite{Newman:1965tw,Janis:1965zz}. Although the original method successfully connects the Schwarzschild and Reissner--Nordström geometries with their rotating counterparts, its complexification step is not unique when the seed metric contains additional matter fields or nonstandard radial functions. For this reason, several modified prescriptions have been proposed. In particular, the noncomplexification formulation introduced by Azreg--Aïnou replaces the static radial functions by real functions $F(r,a,\theta)$ and $H(r,a,\theta)$, allowing one to generate rotating geometries without explicitly complexifying the radial coordinate \cite{AzregAinou:2014pra,AzregAinou:2014aqa}. This corrected formulation is particularly useful for imperfect fluids and effective matter distributions, and it has been employed in different contexts involving regular black holes, modified gravity models, loop--inspired black holes, and other axisymmetric compact objects \cite{Bambi:2013ufa,Ghosh:2015ovj,Kumar:2020owy,Brahma:2020eos,Islam:2022wck,AraujoFilho:2024rss,Kumar:2025ueq,Kumar:2025bim}.

In parallel with these developments, the role of dark matter in gravitational systems remains one of the most important open problems in modern cosmology and astrophysics. According to the standard cosmological model, dark matter accounts for a significant fraction of the total energy budget of the Universe, although its microscopic composition is still unknown \cite{Bertone:2004pz,Freese:2008cz,Wechsler:2018pic,deSwart:2017heh,Planck:2018vyg,Arbey:2021gdg}. Its gravitational effects are supported by several observations, including galactic rotation curves, galaxy clusters, gravitational lensing, large--scale structure formation, and cosmic microwave background measurements \cite{Cebrian:2022brv,Misiaszek:2023sxe}. 

A convenient way to model the gravitational influence of dark matter is to introduce an effective halo density profile around the compact object. In this approach, the matter distribution modifies the energy--momentum tensor and changes the gravitational potential felt by test particles and radiation. Several profiles have been proposed to describe galactic halos, including the Navarro--Frenk--White, Einasto, Burkert, Dehnen, Brownstein, Moore, and Hernquist profiles \cite{Tormen:1996fc,Navarro:1994hi,Navarro:1995iw,Einasto:1965czb,Dutton:2014xda,Graham:2005xx,Burkert:1995yz,Salucci:2000ps,10.1093/mnras/265.1.250,Brownstein:2009gz,Moore:1999gc,Hernquist:1990be}.  

Black holes immersed in dark matter halos have attracted increasing attention in recent years. The surrounding matter distribution can shift the horizon radius, modify the stationary limit surface, change the photon region, affect the shadow size, alter accretion properties, and influence thermodynamic quantities such as temperature, entropy, and heat capacity. These effects have been explored in different black hole models and for several halo profiles, with applications to shadows, lensing, accretion disks, quasinormal modes, greybody factors, and thermal radiation \cite{Al-Badawi:2024asn,Jha:2025xjf,Anjum:2023darkmatter,Liu:2022chb,UktamjonUktamov:2025qts,AHMED2026102368,Lobo:2025kzb,Lobo:2025qap,Mehmood:2025mrg,Zare:2026lgt,Hassanabadi:2026wku,Nieto:2025apz,Lutfuoglu:2026boa,Hamil:2025pte,Al-Badawi:2026kkw,Ahmed:2026gvw,Belchior:2026vwf,Al-Badawi:2025ipr,Konoplya:2021ube,Konoplya:2022hbl,Konoplya:2025ect,Konoplya:2025nqv}. 

Another important aspect concerns particle creation in black hole backgrounds. In static and spherically symmetric spacetimes, Hawking radiation can be obtained by several equivalent methods, including quantum field theory in curved spacetime, Euclidean techniques, gravitational anomaly cancellation, and semiclassical tunneling. In the tunneling picture, the radial part of the classical action develops an imaginary contribution when the particle trajectory crosses the horizon, leading to a thermal emission probability governed by the Hawking temperature \cite{Parikh:1999mf,Srinivasan:1998ty,Shankaranarayanan:2000qv,Angheben:2005rm,Robinson:2005pd,Iso:2006ut}. The Hamilton--Jacobi method provides a particularly direct way to implement this idea, since it relates the pole at the horizon to the surface gravity and reproduces the standard Boltzmann factor. This approach has also been extended to fermions and to charged or rotating black holes \cite{Kerner:2007rr,Kerner:2008qv,Jiang:2005ba,AraujoFilho:2025zzf,AraujoFilho:2024ctw,AraujoFilho:2025jcu,da2010kerr}.

For stationary and axisymmetric black holes, the particle creation process acquires an additional feature associated with the angular velocity of the horizon. The relevant horizon generator is $\chi=\partial_t+\Omega_h\partial_\varphi$, so that the energy entering the thermal spectrum is not simply $\omega$, but the effective combination $\omega-m\Omega_h$ for modes with azimuthal number $m$. Accordingly, the bosonic occupation number takes the form
\[
\langle N_{\omega m}\rangle =
\frac{1}{\exp\left[(\omega-m\Omega_h)/T\right]-1},
\]
up to greybody corrections. This structure reduces to the static result when $\Omega_h\rightarrow0$ and also identifies the superradiant sector $\omega<m\Omega_h$, where bosonic modes may be amplified by extracting rotational energy from the black hole \cite{Starobinsky:1973aij,Bekenstein:1973mi,Brito:2015oca}. In this view, when a static black hole surrounded by a dark matter halo is promoted to an axisymmetric geometry, the particle creation spectrum must encode both the halo--induced modification of the horizon temperature and the rotational chemical potential associated with $\Omega_h$.

Motivated by these points, in this work we investigate a rotating black hole immersed in a Hernquist dark matter halo. Starting from a static black hole surrounded by a Hernquist distribution, we construct the rotating counterpart by employing the noncomplexification formulation of the Newman--Janis algorithm. We then analyze the horizon structure, the stationary limit surfaces, the ergoregion, and the frame--dragging effects induced by the parameters $\rho$, {$r_{\text{s}}$} and $a$. After that, we derive the thermodynamic quantities of the system, including the surface gravity, Hawking temperature, entropy, and heat capacity. We also study the quantum emission process through the Hamilton--Jacobi tunneling method and obtain the corresponding occupation number and particle creation density. Finally, we estimate the Hawking luminosity, evaporation time, and spectral emission rates in suitable analytical regimes.


\section{The rotating black hole solution }

We start from the static and spherically symmetric black hole within the context of Hernquist dark matter halo recently obtained in Ref.~\cite{Jha:2025xjf}, whose line element is written as
\ie
\label{maaaaianametric}
\mathrm{d}s^{2} = -f(r)\mathrm{d}t^{2} + \frac{\mathrm{d}r^{2}}{f(r)} + r^{2}\mathrm{d}\Omega^{2},
\fe
where
\ie
f(r)
=
1-\frac{2M}{r}
-\frac{4\pi\rho_{\mathrm{s}}r_{\mathrm{s}}^{3}}
{r+r_{\mathrm{s}}}.
\fe
Here, $M$ denotes the black hole mass, while $\rho_{\mathrm{s}}$ and $r_{\mathrm{s}}$ characterize the surrounding matter distribution.

In order to construct the rotating counterpart of Eq.~(\ref{maaaaianametric}), we employ the noncomplexification formulation of the Newman--Janis algorithm \cite{n57,n57i,n58,n59,AraujoFilho:2024rss}. Contrary to the original procedure, this prescription does not require an explicit complexification of the radial functions. Instead, it introduces real functions depending on $r$, the rotation parameter $a$, and the polar coordinate $\theta$. This approach has been successfully applied to several static geometries, particularly when the corresponding matter sector cannot be described by a perfect fluid \cite{n60,n65,n62,n61,n63,n64}.

We first express the seed geometry in {retarded} Eddington--Finkelstein coordinates. For this purpose, we introduce the null coordinate $u$ through
\ie
\mathrm{d}u
=
\mathrm{d}t-\frac{\mathrm{d}r}{f(r)}.
\fe
The line element then assumes the form
\ie
\mathrm{d}s^{2}
=
-f(r)\mathrm{d}u^{2}
-2\mathrm{d}u\,\mathrm{d}r
+r^{2}\left(
\mathrm{d}\theta^{2}
+\sin^{2}\theta\,\mathrm{d}\phi^{2}
\right).
\fe

At this stage, we decompose the inverse metric in terms of the null tetrad $Z_{a}^{\mu}=(l^{\mu},n^{\mu},m^{\mu},\bar{m}^{\mu})$ according to
\ie
g^{\mu\nu}
=
-l^{\mu}n^{\nu}
-l^{\nu}n^{\mu}
+m^{\mu}\bar{m}^{\nu}
+m^{\nu}\bar{m}^{\mu}.
\fe
A convenient tetrad basis associated with the above metric is
\ie
\begin{split}
l^{\mu}
&=
\delta_{r}^{\mu},
\\
n^{\mu}
&=
\delta_{u}^{\mu}
-\frac{f(r)}{2}\delta_{r}^{\mu},
\\
m^{\mu}
&=
\frac{1}{\sqrt{2}\,r}
\left(
\delta_{\theta}^{\mu}
+\frac{i}{\sin\theta}\delta_{\phi}^{\mu}
\right).
\end{split}
\fe
Here, $\bar{m}^{\mu}$ denotes the complex conjugate of $m^{\mu}$. These vectors satisfy
\ie
l^{\mu}l_{\mu}
=
n^{\mu}n_{\mu}
=
m^{\mu}m_{\mu}
=
l^{\mu}m_{\mu}
=
n^{\mu}m_{\mu}
=
0,
\fe
together with
\ie
l^{\mu}n_{\mu}
=
-m^{\mu}\bar{m}_{\mu}
=
-1.
\fe

We now introduce the rotation parameter $a$ by applying the Newman--Janis transformation to the tetrad basis. Within the Azreg--A\"{\i}nou prescription \cite{azreg2014generating,afrim}, the quantities $f(r)$ and $r^{2}$ are replaced, respectively, by two real functions $F(r,a,\theta)$ and $H(r,a,\theta)$. The transformed tetrad can therefore be expressed as
\ie
\begin{split}
l'^{\mu}
&=
\delta_{r}^{\mu},
\\
n'^{\mu}
&=
\delta_{u}^{\mu}
-\frac{F(r,a,\theta)}{2}\delta_{r}^{\mu},
\\
m'^{\mu}
&=
\frac{1}{\sqrt{2H(r,a,\theta)}}
\left[
ia\sin\theta
\left(
\delta_{u}^{\mu}-\delta_{r}^{\mu}
\right)
+\delta_{\theta}^{\mu}
+\frac{i}{\sin\theta}\delta_{\phi}^{\mu}
\right].
\end{split}
\fe
The transformed inverse metric follows directly from
\ie
g^{\mu\nu}
=
-l'^{\mu}n'^{\nu}
-l'^{\nu}n'^{\mu}
+m'^{\mu}\bar{m}'^{\nu}
+m'^{\nu}\bar{m}'^{\mu}.
\fe

By inverting this expression, we obtain the rotating geometry in Eddington--Finkelstein coordinates:
\ie
\begin{split}
\mathrm{d}s^{2}
={}&
-F(r,a,\theta)\mathrm{d}u^{2}
-2\mathrm{d}u\,\mathrm{d}r
+2a\sin^{2}\theta
\left[
F(r,a,\theta)-1
\right]
\mathrm{d}u\,\mathrm{d}\phi
\\
&+
2a\sin^{2}\theta\,
\mathrm{d}r\,\mathrm{d}\phi
+
H(r,a,\theta)\mathrm{d}\theta^{2}
\\
&+
\sin^{2}\theta
\left[
H(r,a,\theta)
+a^{2}\sin^{2}\theta
\left(
2-F(r,a,\theta)
\right)
\right]
\mathrm{d}\phi^{2}.
\end{split}
\fe
Notice that the static line element is recovered when $a\rightarrow0$, provided that $F(r,0,\theta)=f(r)$ and $H(r,0,\theta)=r^{2}$.
To express the metric in Boyer--Lindquist coordinates, we perform the transformations
\ie
\mathrm{d}u = \mathrm{d}t+\lambda(r)\mathrm{d}r, \qquad \mathrm{d}\phi = \mathrm{d}\varphi+\Tilde{\rho}(r)\mathrm{d}r.
\fe
The functions $\lambda(r)$ and $\Tilde{\rho}(r)$ must depend only on the radial coordinate. This requirement is satisfied by choosing
\ie
H(r,a,\theta) = \Sigma \equiv r^{2}+a^{2}\cos^{2}\theta
\fe
and
\ie
F(r,a,\theta) = \frac{r^{2}f(r)+a^{2}\cos^{2}\theta}{\Sigma}.
\fe
The first relation also guarantees the vanishing of the mixed Einstein tensor component $G_{r\theta}$ \cite{azreg2014generating,afrim}. Defining
\ie
\Delta(r) = r^{2}f(r)+a^{2},
\fe
we can equivalently write
\ie
F(r,a,\theta) = \frac{\Delta(r)-a^{2}\sin^{2}\theta}{\Sigma}.
\fe
The radial functions entering the Boyer--Lindquist transformation are consequently given by
\ie
\lambda(r) = -\frac{r^{2}+a^{2}}{\Delta(r)},
\qquad \Tilde{\rho}(r) = -\frac{a}{\Delta(r)}.
\fe

After substituting these expressions and relabeling $\varphi\rightarrow\phi$, we arrive at the rotating extension of the {Hernquist} black hole:
\ie
\begin{split}
\label{newbumblebeerotating}
\mathrm{d}s^{2} ={}& -\frac{\Delta(r)-a^{2}\sin^{2}\theta}{\Sigma} \,\mathrm{d}t^{2} + \frac{\Sigma}{\Delta(r)} \,\mathrm{d}r^{2} + \Sigma\,\mathrm{d}\theta^{2} \\ &- 2a\sin^{2}\theta \left[ 1- \frac{\Delta(r)-a^{2}\sin^{2}\theta}{\Sigma} \right] \mathrm{d}t\,\mathrm{d}\phi \\ &+ \frac{\sin^{2}\theta}{\Sigma} \left[ \left(r^{2}+a^{2}\right)^{2} -a^{2}\Delta(r)\sin^{2}\theta \right] \mathrm{d}\phi^{2},
\end{split}
\fe
where
\ie
\Sigma = r^{2}+a^{2}\cos^{2}\theta, \qquad \Delta(r) = r^{2}+a^{2} -2Mr -\frac{4\pi\rho_{\mathrm{s}}r_{\mathrm{s}}^{3}r^{2}} {r+r_{\mathrm{s}}}.
\fe
Equation~(\ref{newbumblebeerotating}) consistently reduces to the original static geometry when $a\rightarrow 0$. In addition, the Kerr spacetime is recovered after removing the surrounding matter contribution, namely, in the limit $\rho_{\mathrm{s}}\rightarrow0$. The possible horizons are determined by the roots of $\Delta(r)=0$. At this stage, for completeness, we analyze the radial behavior of $\Delta(r)$ in Fig.~\ref{deltaDDD}. In the left panel, we fix $a=0.1$ and $M=1$ and vary $\rho$, whereas, in the right panel, we set $\rho=0.1$ and $M=1$ and vary $a$. In both panels, the corresponding configurations are displayed as functions of the radial coordinate $r$.


\section{Horizon geometry and rotational effects}


\subsection{Horizons, stationary limit surfaces, and ergoregions}

The horizons of the rotating black hole immersed in the Hernquist dark matter distribution are determined by the roots of $\Delta(r)=0$. In the present case, the relevant metric functions are
\begin{equation}
{ 
f(r) = 1-\frac{2M}{r} -\frac{4\pi\rho r_{\text{s}}^{3}}{r+r_{\text{s}}}, \qquad \Delta(r)=a^{2}+r^{2}f(r) =r^{2}+a^{2}-2Mr -\frac{4\pi\rho r_{\text{s}}^{3}r^{2}}{r+r_{\text{s}}}.}
\label{generalDelta}
\end{equation}
{
Here, $\rho\equiv\rho_{\text{s}}$ is the characteristic density and $r_{\text{s}}$ is the scale radius of the Hernquist halo. In particular, $r_{\text{s}}$ is an independent environmental parameter and must not be identified with the Schwarzschild radius $2M$. It is useful to introduce the dimensionless quantities}
\begin{equation}
{
\bar r=\frac{r}{M}, \qquad \bar a=\frac{a}{M}, \qquad \bar r_{\text{s}}=\frac{r_{\text{s}}}{M}, \qquad \bar\rho=\rho M^{2},}
\end{equation}
{ in terms of which}
\begin{equation}
{
\frac{\Delta}{M^{2}} = \bar r^{2}+\bar a^{2}-2\bar r -\frac{4\pi\bar\rho\,\bar r_{\text{s}}^{3}\bar r^{2}} {\bar r+\bar r_{\text{s}}}.}
\label{dimensionlessDelta}
\end{equation}
{
The rotating geometry is characterized by the three independent dimensionless parameters $\bar a$, $\bar\rho$, and $\bar r_{\text{s}}$. This distinction is also observationally relevant. For the static seed and $\rho M^{2}=1$, Ref.~\cite{Jha:2025xjf} reported the $1\sigma$ ($2\sigma$) upper bounds $r_{\text{s}}/M\leq0.302507$ ($0.38625$) from M87$^{*}$/EHT, $r_{\text{s}}/M\leq0.203867$ ($0.289123$) from Sgr A$^{*}$/Keck, and $r_{\text{s}}/M\leq0.118254$ ($0.257937$) from Sgr A$^{*}$/VLTI. It is important to notice that these conditional bounds were obtained for a nonrotating configuration and therefore cannot be transferred directly to the rotating solution; nevertheless, they show why $r_{\text{s}}$ should be retained as an independent parameter.}

Accordingly, the horizon equation can be written as
\begin{equation}
{
r^{2}-2Mr+a^{2} -\frac{4\pi\rho r_{\text{s}}^{3}r^{2}}{r+r_{\text{s}}}=0.}
\end{equation}
After multiplying by {$(r+r_{\text{s}})$}, we obtain the cubic equation
\begin{equation}
{
F(r,\rho,r_{\text{s}}) \equiv (r^{2}-2Mr+a^{2})(r+r_{\text{s}}) -4\pi\rho r_{\text{s}}^{3}r^{2}=0,}
\label{Fdef}
\end{equation}
or, similarly,
\begin{equation}
{
F(r,\rho,r_{\text{s}}) =r^{3} +\big(r_{\text{s}}-2M-4\pi\rho r_{\text{s}}^{3}\big)r^{2} +\big(a^{2}-2Mr_{\text{s}}\big)r +a^{2}r_{\text{s}}.}
\end{equation}

For $\rho=0$, Eq.~\eqref{Fdef} reduces to
\begin{equation}
{
F(r,0,r_{\text{s}}) =(r^{2}-2Mr+a^{2})(r+r_{\text{s}})=0.}
\end{equation}
The root {$r=-r_{\text{s}}$} results from multiplying the original equation by {$(r+r_{\text{s}})$} and does not represent a physical horizon. The remaining roots correspond to the Kerr horizons,
\begin{equation}
r_{\pm}^{(0)} = M\pm\sqrt{M^{2}-a^{2}}.
\end{equation}

Since the cubic equation does not provide a sufficiently compact expression for the physical roots, we consider the perturbative regime in which $\rho$ is small. The deformed horizons can then be expressed as
\begin{equation}
r_{\pm} = r_{\pm}^{(0)} +\rho\,r_{\pm}^{(1)} +\mathcal{O}(\rho^{2}).
\label{horizonexpansion}
\end{equation}
Substituting Eq.~\eqref{horizonexpansion} into {$F(r_{\pm},\rho,r_{\text{s}})=0$} and expanding up to first order in $\rho$, we obtain
\begin{equation}
{
0 = \left. \frac{\partial F}{\partial r} \right|_{(r_{\pm}^{(0)},0,r_{\text{s}})} \rho\,r_{\pm}^{(1)} + \left. \frac{\partial F}{\partial \rho} \right|_{(r_{\pm}^{(0)},0,r_{\text{s}})} \rho +\mathcal{O}(\rho^{2}).}
\end{equation}
The first--order correction is therefore given by
\begin{equation}
{
r_{\pm}^{(1)} = -\left. \frac{\partial_{\rho}F}{\partial_{r}F} \right|_{(r_{\pm}^{(0)},0,r_{\text{s}})}.}
\label{deltar}
\end{equation}
From Eq.~\eqref{Fdef}, we find
\begin{equation}
{
\frac{\partial F}{\partial\rho} =-4\pi r_{\text{s}}^{3}r^{2},}
\end{equation}
whereas
\begin{equation}
{
\left. \frac{\partial F}{\partial r} \right|_{\rho=0} = \frac{\mathrm{d}}{\mathrm{d}r} \left[ (r^{2}-2Mr+a^{2})(r+r_{\text{s}}) \right].}
\end{equation}
Using
\begin{equation}
{
\left. \frac{\partial F}{\partial r} \right|_{(r_{\pm}^{(0)},0,r_{\text{s}})} = 2\big(r_{\pm}^{(0)}-M\big) \big(r_{\pm}^{(0)}+r_{\text{s}}\big),}
\end{equation}
the two horizon radii become
\begin{equation}
{
r_{\pm} = r_{\pm}^{(0)} + \frac{ 2\pi r_{\text{s}}^{3}\big(r_{\pm}^{(0)}\big)^{2} }{ \big(r_{\pm}^{(0)}-M\big) \big(r_{\pm}^{(0)}+r_{\text{s}}\big) } \,\rho +\mathcal{O}(\rho^{2}).}
\label{generalhorizons}
\end{equation}

In particular, the outer event horizon radius is
\begin{equation}
{
\begin{aligned}
r_{h} ={}& M+\sqrt{M^{2}-a^{2}} + \frac{ 2\pi r_{\text{s}}^{3} \big(M+\sqrt{M^{2}-a^{2}}\big)^{2} }{ \sqrt{M^{2}-a^{2}} \big(M+\sqrt{M^{2}-a^{2}}+r_{\text{s}}\big) } \,\rho +\mathcal{O}(\rho^{2}).
\end{aligned}}
\label{eventhooo}
\end{equation}
{
For positive $\rho$ and $r_{\text{s}}$, the first--order correction is positive and the Hernquist halo displaces the outer horizon toward larger radial coordinates. The separate role of $r_{\text{s}}$ follows directly from the factor $r_{\text{s}}^{3}/(r_{+}^{(0)}+r_{\text{s}})$: the correction grows cubically for $r_{\text{s}}\ll r_{+}^{(0)}$ and quadratically for $r_{\text{s}}\gg r_{+}^{(0)}$.}

{
The monotonic dependence on both halo parameters can also be established without expanding the horizon radius. Implicit differentiation of $\Delta(r_{h})=0$ gives}
\begin{equation}
{
\frac{\partial r_{h}}{\partial\rho} = \frac{ 4\pi r_{\text{s}}^{3}r_{h}^{2} }{ (r_{h}+r_{\text{s}})\Delta'(r_{h}) }, \qquad \frac{\partial r_{h}}{\partial r_{\text{s}}} = \frac{ 4\pi\rho r_{\text{s}}^{2}r_{h}^{2} (3r_{h}+2r_{\text{s}}) }{ (r_{h}+r_{\text{s}})^{2}\Delta'(r_{h}) }.}
\label{horizonmonotonicity}
\end{equation}
{
Both derivatives are positive for a nonextremal outer horizon, for which $\Delta'(r_{h})>0$. Equation~\eqref{horizonmonotonicity} is exact within the leading--order Hernquist lapse adopted in Eq.~\eqref{generalDelta}.}

The perturbative result in Eq.~\eqref{eventhooo} assumes a nonextremal Kerr seed, $M^{2}>a^{2}$, together with the condition that the correction proportional to $\rho$ remains small compared with the zeroth--order horizon radius. {Similarly, the dimensionless Hernquist contribution $4\pi\rho r_{\text{s}}^{3}/(r+r_{\text{s}})$ must remain perturbative in the radial domain under consideration.} In particular, Eq.~\eqref{eventhooo} cannot be directly applied to the extremal limit $a\to M$, since its first--order correction becomes singular in this limit. The extremal configuration must instead be obtained directly from the simultaneous conditions
\begin{equation}
{
F(r,\rho,r_{\text{s}})=0, \qquad \frac{\partial F(r,\rho,r_{\text{s}})}{\partial r}=0.}
\end{equation}
Expanding these conditions for small $\rho$ gives
\begin{equation}
{
r_{\mathrm{ext}} = M+ \frac{ 2\pi r_{\text{s}}^{3}M(M+2r_{\text{s}}) }{ (M+r_{\text{s}})^{2} }\rho +\mathcal{O}(\rho^{2}), \qquad a_{\mathrm{ext}}^{2} = M^{2} + \frac{ 4\pi r_{\text{s}}^{3}M^{2} }{ M+r_{\text{s}} }\rho +\mathcal{O}(\rho^{2}).}
\label{generalextremal}
\end{equation}

It is also important to examine the nonrotating limit. Setting $a=0$ in the exact horizon equation gives
\begin{equation}
{
r_{h} = \frac{1}{2} \left[ 2M+4\pi\rho r_{\text{s}}^{3}-r_{\text{s}} + \sqrt{ \big(r_{\text{s}}-2M-4\pi\rho r_{\text{s}}^{3}\big)^{2} +8Mr_{\text{s}} } \right],}
\label{exactstatichorizon}
\end{equation}
whose perturbative expansion reads
\begin{equation}
{
r_{h} = 2M + \frac{ 8\pi M r_{\text{s}}^{3} }{ 2M+r_{\text{s}} }\rho +\mathcal{O}(\rho^{2}).
}
\label{staticlimit}
\end{equation}
{
In other words, the Schwarzschild value $r_{h}=2M$ is recovered when the halo contribution is removed, either through $\rho\to0$ or $r_{\text{s}}\to0$. Equation~\eqref{exactstatichorizon} also reproduces the static horizon obtained in Ref.~\cite{Jha:2025xjf}.}

\begin{figure}
    \centering
    \includegraphics[scale=0.475]{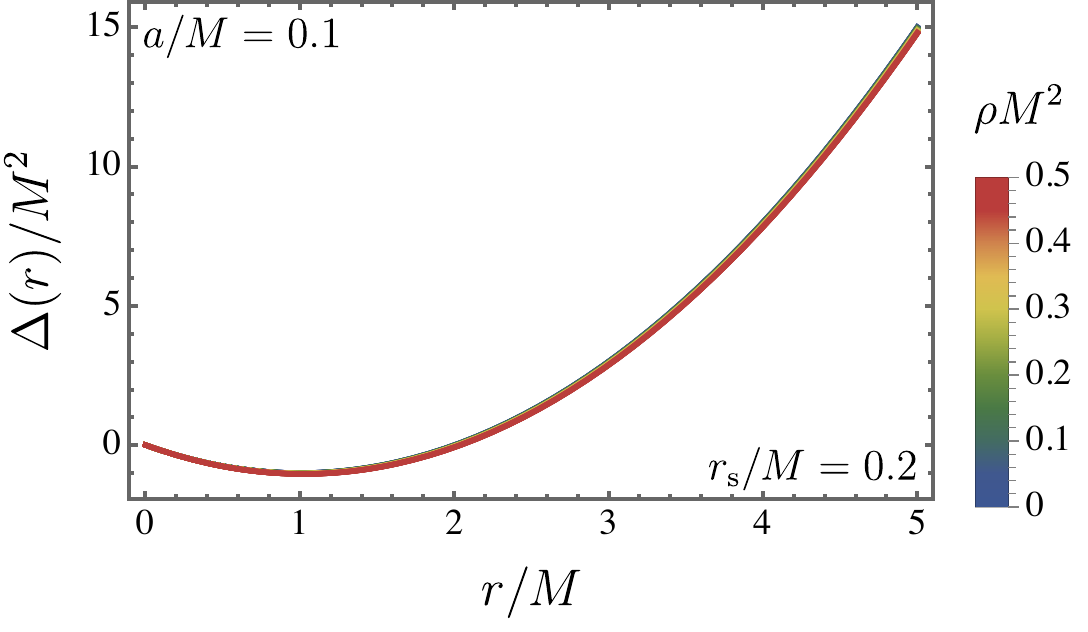}
    \includegraphics[scale=0.475]{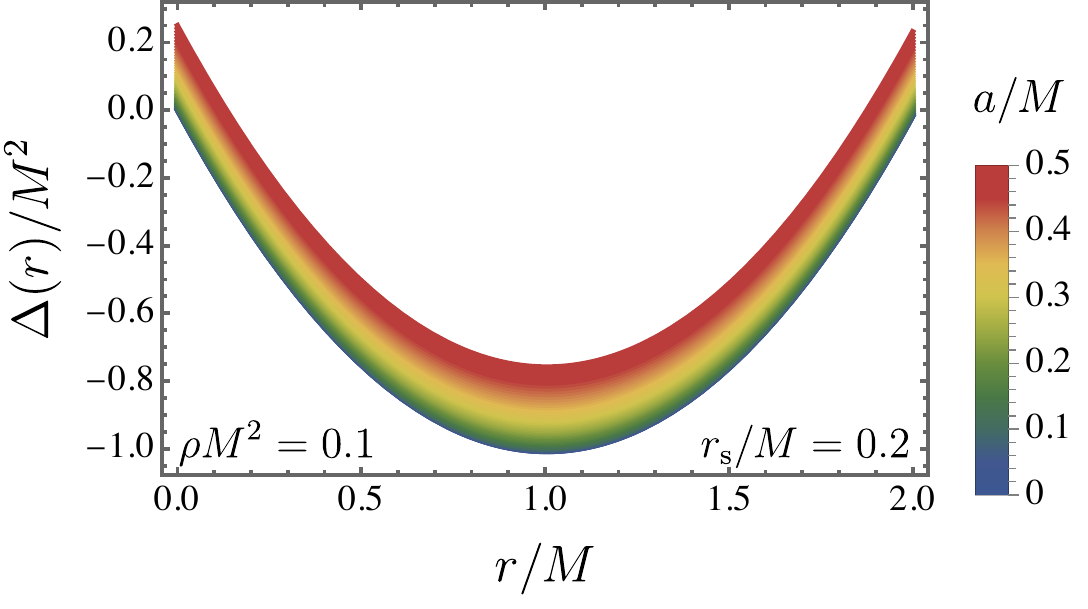}
    \caption{{Radial behavior of $\Delta(r)$ for representative values of the model parameters at a fixed halo scale radius $r_{\text{s}}$. In the left panel, we fix $a=0.1$ and $M=1$ and vary the characteristic density $\rho$, whereas, in the right panel, we set $\rho=0.1$ and $M=1$ and vary the rotation parameter $a$ (for $r_{\text{s}}/M =0.2$).}}
    \label{deltaDDD}
\end{figure}

\begin{figure}
    \centering
    \includegraphics[scale=0.475]{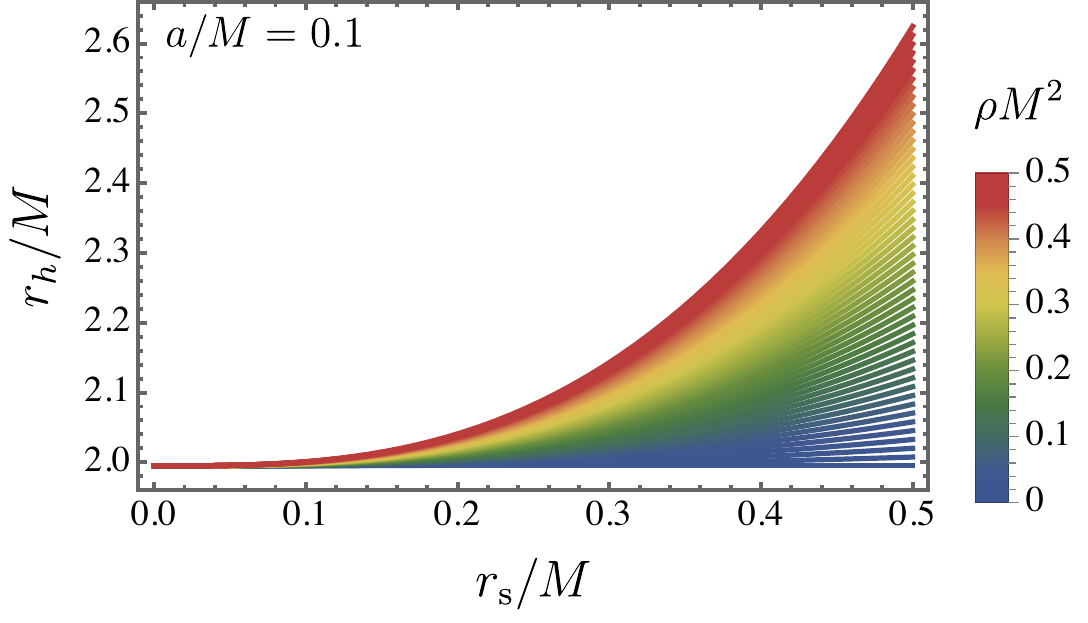}
    \includegraphics[scale=0.475]{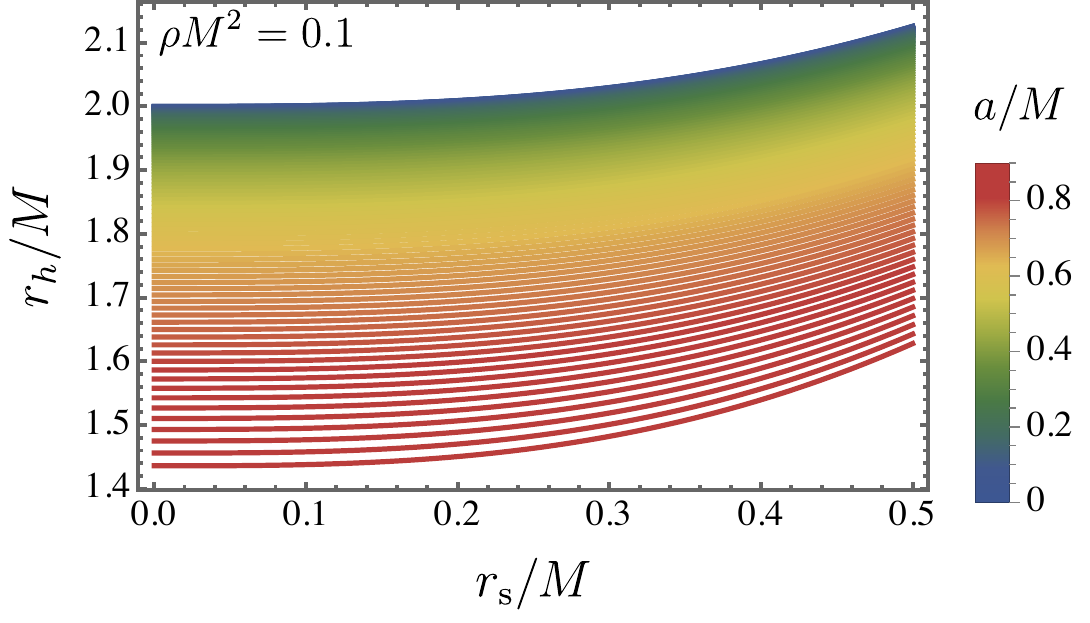}
    \caption{{Behavior of the dimensionless outer event horizon radius $r_{h}/M$ for different values of the characteristic density $\rho M^{2}$ in the left panel and of the rotation parameter $a/M$ in the right panel. The independent scale radius $r_{\text{s}}/M$ and the remaining parameters are fixed at the values indicated in the respective panels.}}
    \label{eventhorizon}
\end{figure}

\begin{figure}
    \centering
    \includegraphics[scale=0.52]{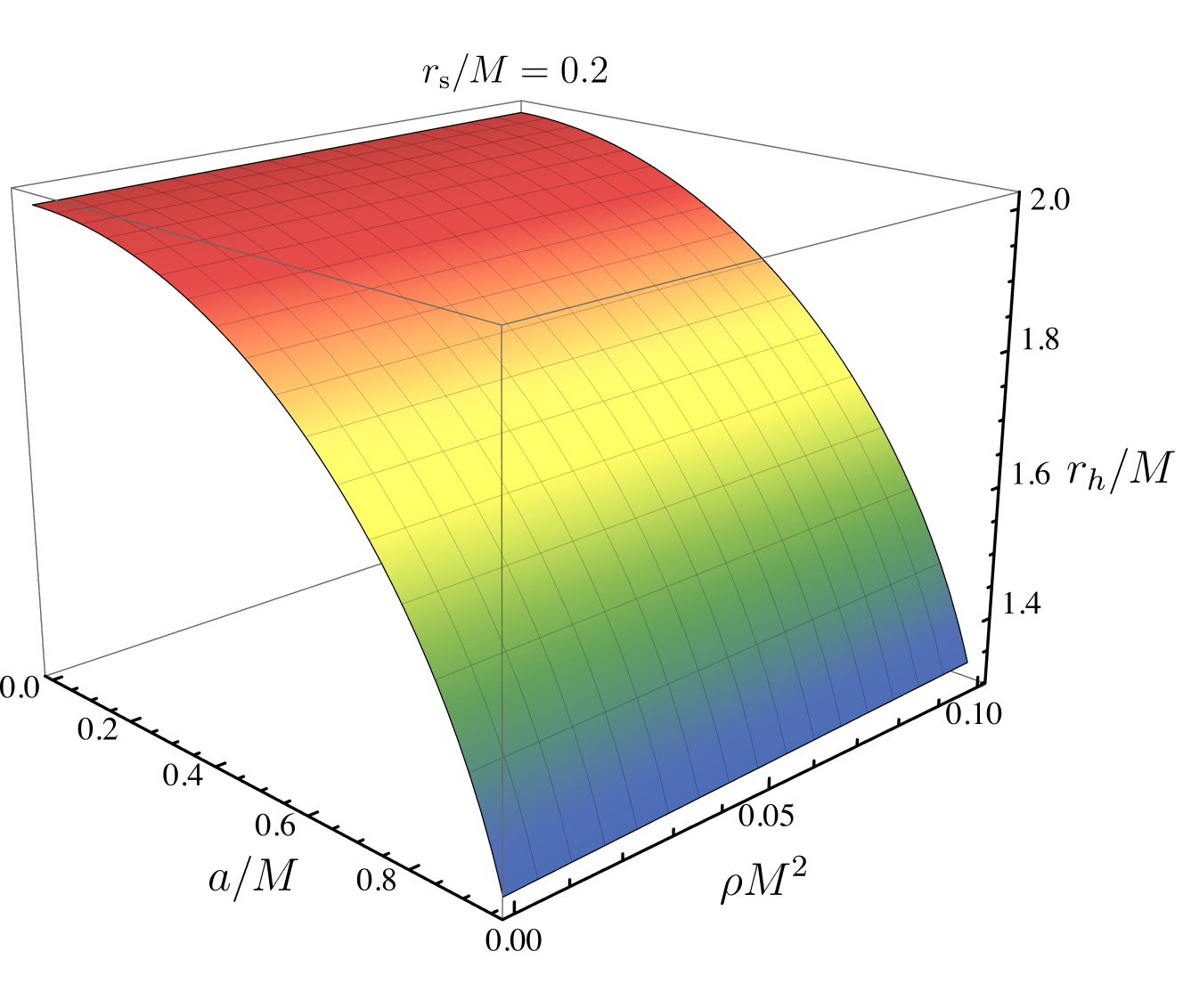}
    \caption{{Three--dimensional representation of the dimensionless outer event horizon radius $r_{h}/M$ in the $(a/M,\rho M^{2})$ parameter space for a fixed value of the independent halo scale radius $r_{\text{s}}/M$.}}
    \label{eventhorizon3}
\end{figure}

\begin{figure}
    \centering
    \includegraphics[scale=0.51]{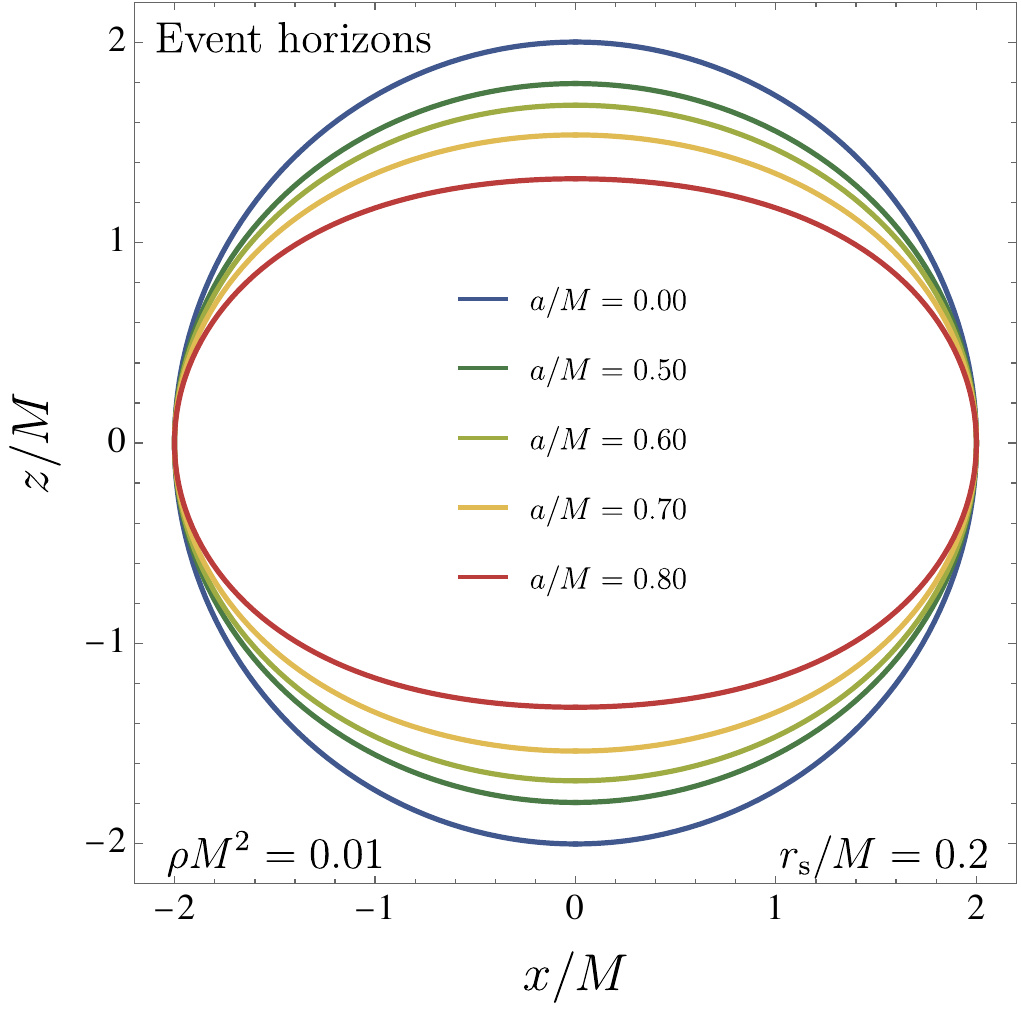}
    \includegraphics[scale=0.51]{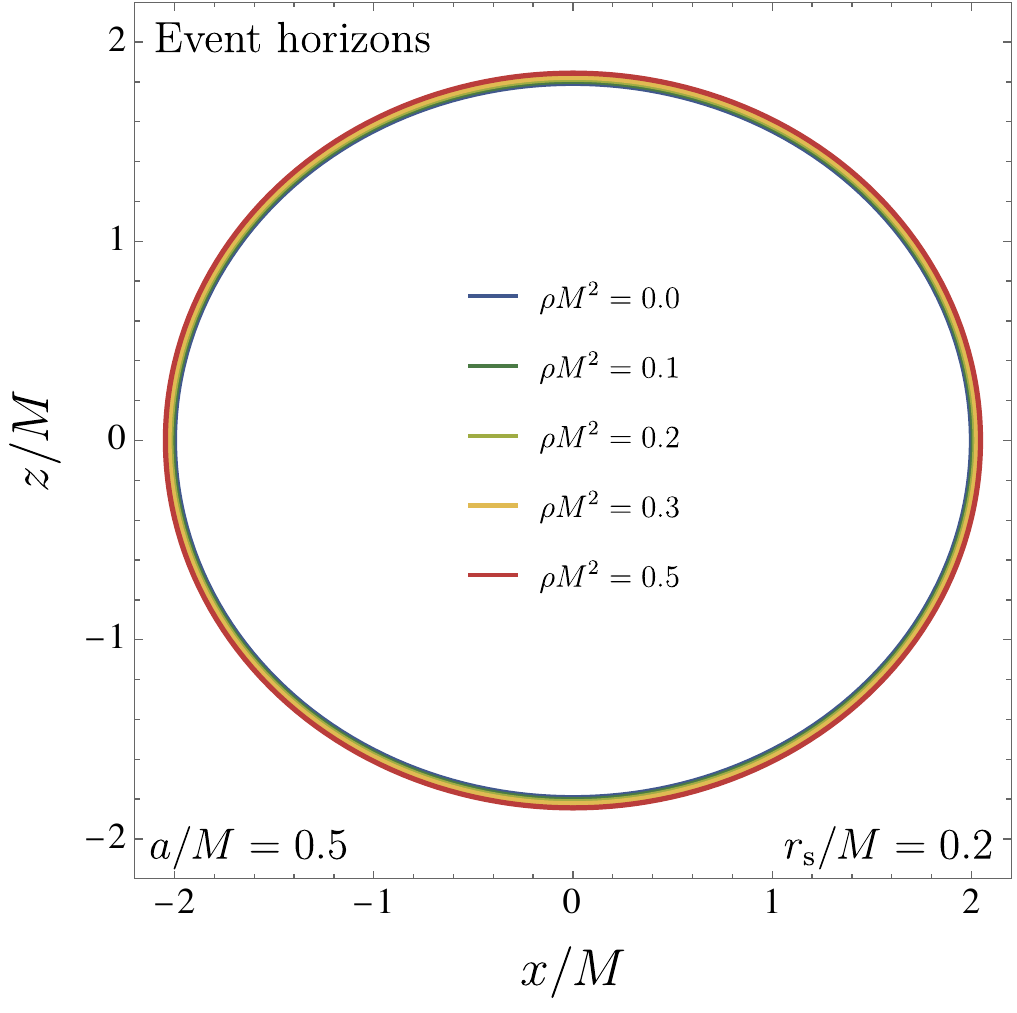}
    \caption{{Two--dimensional representations of the outer event horizon for selected values of the characteristic density, rotation parameter, and halo scale radius. The dimensionless parameter choices are indicated in the respective panels.}}
    \label{eventhorizons}
\end{figure}

{
In Fig.~\ref{deltaDDD}, we display the radial behavior of $\Delta(r)$ for different values of $\rho$ and $a$ at a fixed value of the independent scale radius $r_{\text{s}}$. At fixed $M$, $a$, and $r_{\text{s}}$, increasing $\rho$ enhances the negative dark matter contribution and shifts $\Delta(r)$ downward. At fixed $M$, $a$, and $\rho$, increasing $r_{\text{s}}$ produces the same qualitative displacement because}
\begin{equation}
{
\frac{\partial\Delta}{\partial r_{\text{s}}} = -\frac{ 4\pi\rho r^{2}r_{\text{s}}^{2} (3r+2r_{\text{s}}) }{ (r+r_{\text{s}})^{2} } < 0.}
\label{DeltaRsDerivative}
\end{equation}
{
In this manner, the outer positive root moves toward larger values of $r$ as either halo parameter increases. By contrast, increasing $a$ contributes through the positive term $a^{2}$ and shifts $\Delta(r)$ upward, generally reducing the outer horizon radius within the domain in which the perturbative approximation remains valid. The corresponding behavior of $r_{h}$ is displayed in Figs.~\ref{eventhorizon}--\ref{eventhorizons}. All numerical panels represent dimensionless slices of the $(\bar a,\bar\rho,\bar r_{\text{s}})$ parameter space, and the value of $\bar r_{\text{s}}$ must therefore be specified whenever it is held fixed.}

We now analyze the stationary limit surfaces associated with the rotating configuration. These surfaces are determined by the condition $g_{tt}=0$. For the present metric,
\begin{equation}
g_{tt} = -\frac{\Delta(r)-a^{2}\sin^{2}\theta}{\Sigma(r,\theta)}, \qquad \Sigma(r,\theta)=r^{2}+a^{2}\cos^{2}\theta.
\end{equation}
Since $\Sigma(r,\theta)$ is nonvanishing outside the curvature singularity, the stationary limit surfaces follow from $\Delta(r)-a^{2}\sin^{2}\theta=0$, or, equivalently, $r^{2}f(r)+a^{2}\cos^{2}\theta=0$. After multiplying by {$(r+r_{\text{s}})$}, the corresponding cubic equation becomes
\begin{equation}
{
\mathcal{E}(r,\rho,r_{\text{s}},\theta) \equiv \big(r^{2}-2Mr+a^{2}\cos^{2}\theta\big)(r+r_{\text{s}}) -4\pi\rho r_{\text{s}}^{3}r^{2}=0.}
\label{ergocubic}
\end{equation}

In the absence of the Hernquist contribution, Eq.~\eqref{ergocubic} reproduces the Kerr stationary limit surfaces,
\begin{equation}
r_{e,0}^{\pm}(\theta) = M\pm\sqrt{M^{2}-a^{2}\cos^{2}\theta}.
\end{equation}
Following the same perturbative procedure employed for the horizons, we write
\begin{equation}
r_{e}^{\pm}(\theta) = r_{e,0}^{\pm}(\theta) + \rho\,r_{e,1}^{\pm}(\theta) + \mathcal{O}(\rho^{2}).
\end{equation}
The resulting first-order correction is
\begin{equation}
{
r_{e,1}^{\pm}(\theta) = \frac{ 2\pi r_{\text{s}}^{3} \big[r_{e,0}^{\pm}(\theta)\big]^{2} }{ \big[r_{e,0}^{\pm}(\theta)-M\big] \big[r_{e,0}^{\pm}(\theta)+r_{\text{s}}\big] }. }
\end{equation}

Defining
\begin{equation}
\mathcal{S}(\theta) = \sqrt{M^{2}-a^{2}\cos^{2}\theta},
\end{equation}
the outer stationary limit surface is given by
\begin{equation}
{
r_{e}^{+}(\theta) = M+\mathcal{S}(\theta) + \frac{ 2\pi r_{\text{s}}^{3} \big[M+\mathcal{S}(\theta)\big]^{2} }{ \mathcal{S}(\theta) \big[M+\mathcal{S}(\theta)+r_{\text{s}}\big] } \,\rho +\mathcal{O}(\rho^{2}),}
\label{outerergo}
\end{equation}
whereas the inner stationary limit surface reads
\begin{equation}
{
r_{e}^{-}(\theta) = M-\mathcal{S}(\theta) - \frac{ 2\pi r_{\text{s}}^{3} \big[M-\mathcal{S}(\theta)\big]^{2} }{ \mathcal{S}(\theta) \big[M-\mathcal{S}(\theta)+r_{\text{s}}\big] } \,\rho +\mathcal{O}(\rho^{2}).}
\label{innerergo}
\end{equation}

Unlike the horizons, the stationary limit surfaces depend explicitly on the polar coordinate $\theta$. At the rotation axis, $\theta=0,\pi$, we have $r_{e}^{+}(0)=r_{e}^{+}(\pi)=r_{h}$, since the stationary-limit and horizon equations coincide at the poles. At the equatorial plane, $\theta=\pi/2$, the outer stationary limit surface becomes
\begin{equation}
{
r_{e}^{+}\left(\frac{\pi}{2}\right) = \frac{1}{2} \left[ 2M+4\pi\rho r_{\text{s}}^{3}-r_{\text{s}} + \sqrt{ \big(r_{\text{s}}-2M-4\pi\rho r_{\text{s}}^{3}\big)^{2} +8Mr_{\text{s}} } \right], }
\label{equatorialSLSexact}
\end{equation}
{
which is independent of $a$ exactly within the adopted lapse. Its small $\rho$ expansion is}
\begin{equation}
{
r_{e}^{+}\left(\frac{\pi}{2}\right) = 2M + \frac{ 8\pi M r_{\text{s}}^{3} }{ 2M+r_{\text{s}} }\rho +\mathcal{O}(\rho^{2}).}
\end{equation}
{
The ergoregion is located between $r_{h}$ and $r_{e}^{+}(\theta)$, reaches its largest radial extension at the equatorial plane, and vanishes at the rotation axis. Both bounding surfaces acquire an explicit dependence on $r_{\text{s}}$. At fixed $a$, the first-order outward shift of the event horizon is at least as large as that of the outer stationary limit surface; Then, the coordinate thickness $r_{e}^{+}(\theta)-r_{h}$ decreases with increasing $\rho$ at this order, except at the poles, where it remains zero. In the nonrotating limit, $a\to0$, the stationary limit surface coincides with the event horizon and the ergoregion disappears.}

\begin{figure}
    \centering
     \includegraphics[scale=0.51]{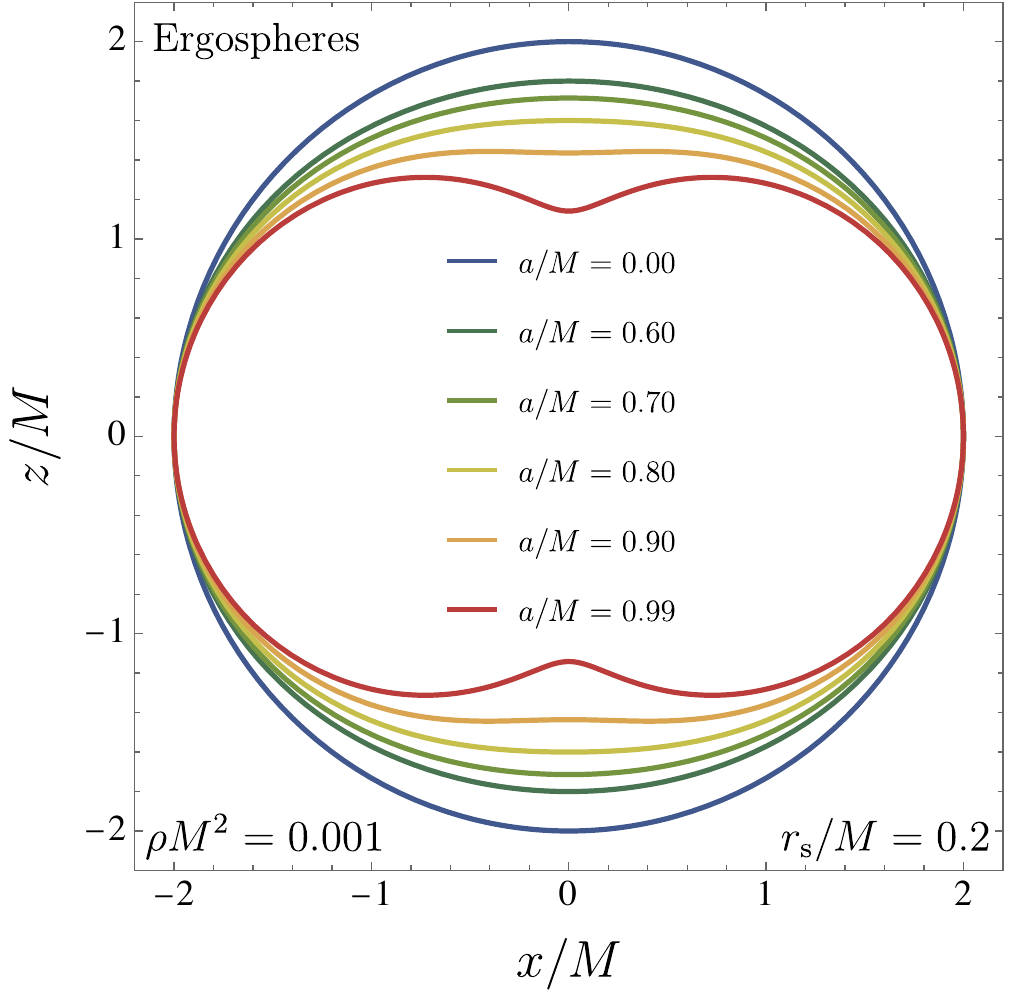}
     \includegraphics[scale=0.52]{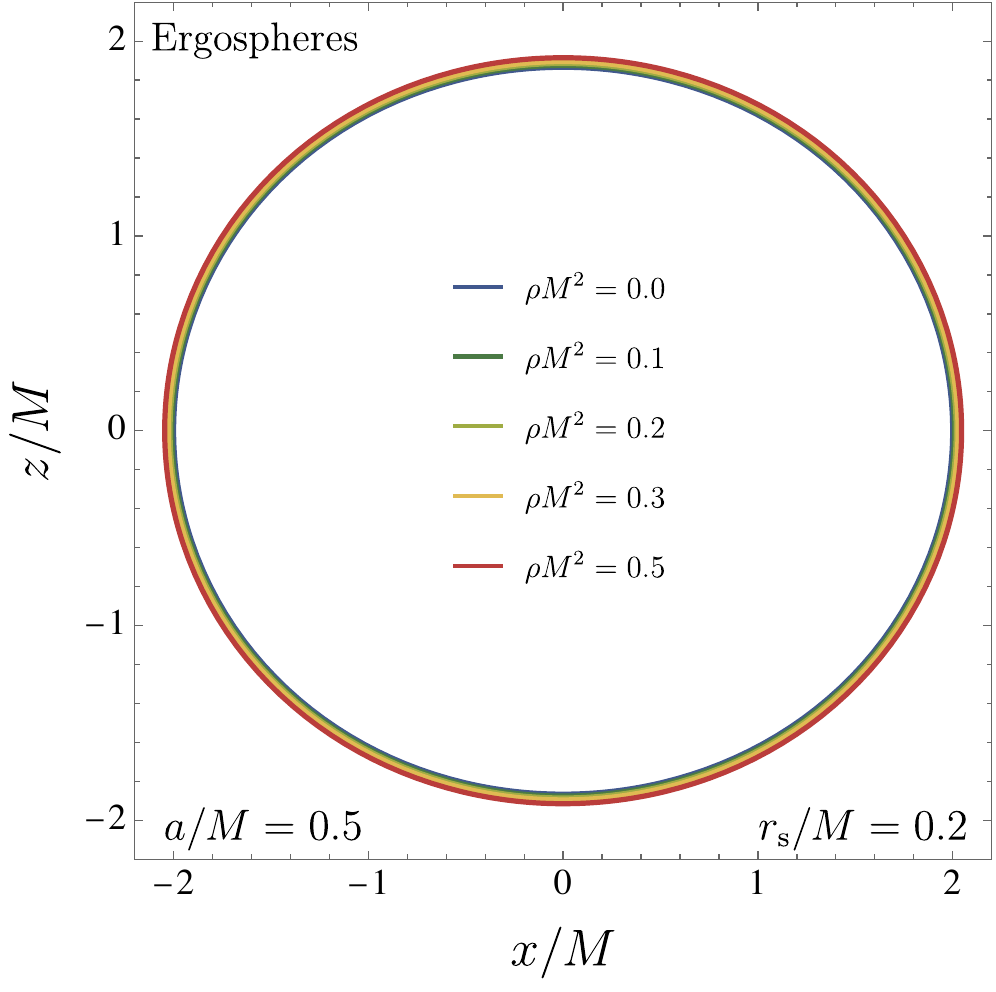}
    \caption{{Ergosphere profiles for different configurations of the system at a fixed halo scale radius $r_{\text{s}}$. In the left panel, we fix $\rho=0.001$ and $M=1$ and vary the rotation parameter $a$, whereas, in the right panel, we set $a=0.5$ and $M=1$ and consider different values of $\rho$. The fixed value of $r_{\text{s}}/M$ is indicated in the respective panels.}}
    \label{ergospheres}
\end{figure}

{
In Fig.~\ref{ergospheres}, we display the ergospheres for different configurations of the system at a fixed halo scale radius $r_{\text{s}}$. In the left panel, we fix $\rho=0.001$ and $M=1$ and vary the rotation parameter $a$, whereas, in the right panel, we set $a=0.5$ and $M=1$ and consider different values of $\rho$. The value of $r_{\text{s}}/M$ is an additional physical input and must be stated in both panels. }

\begin{figure}
    \centering
     \includegraphics[scale=0.4]{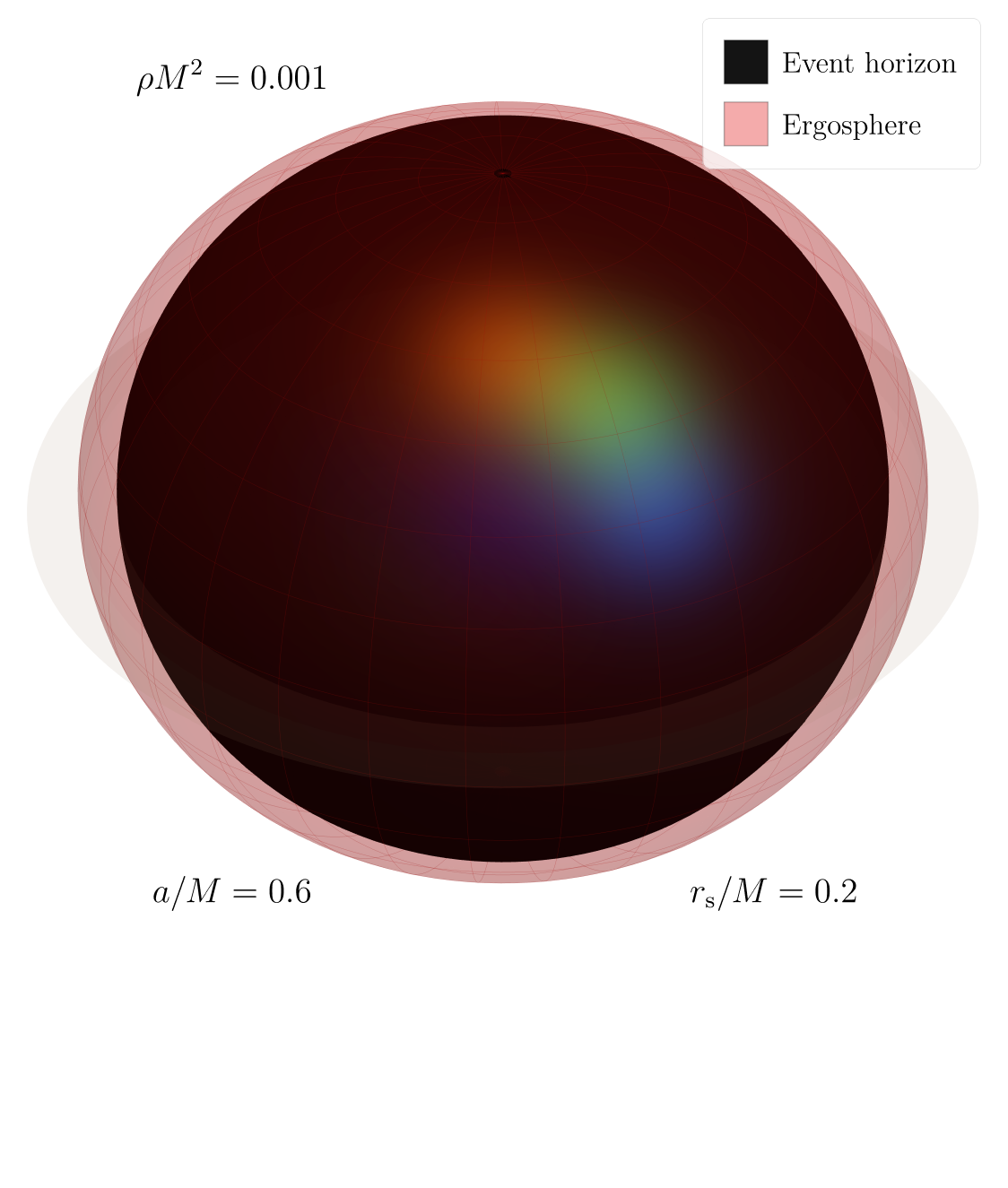}
      \includegraphics[scale=0.4]{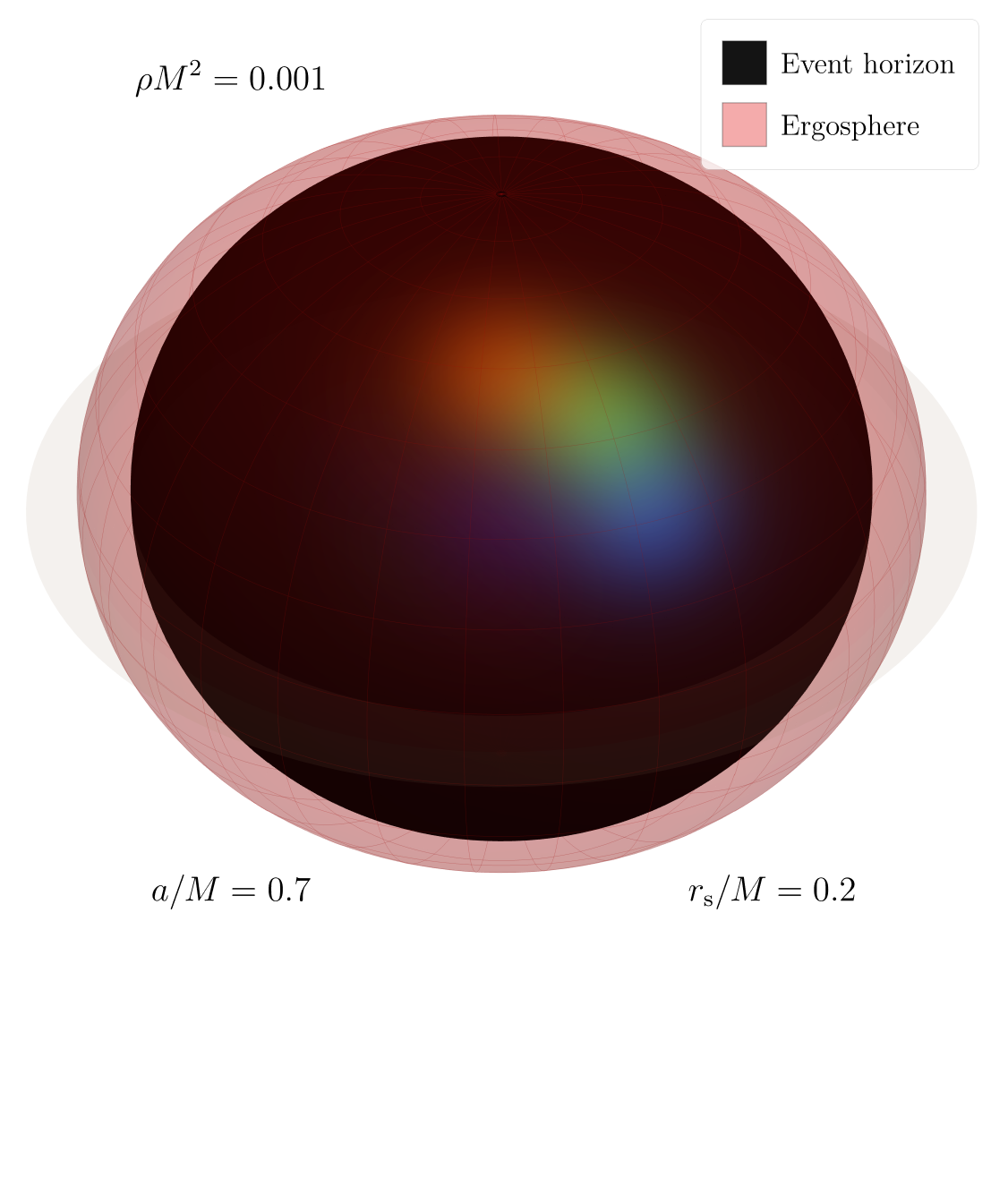}
       \includegraphics[scale=0.4]{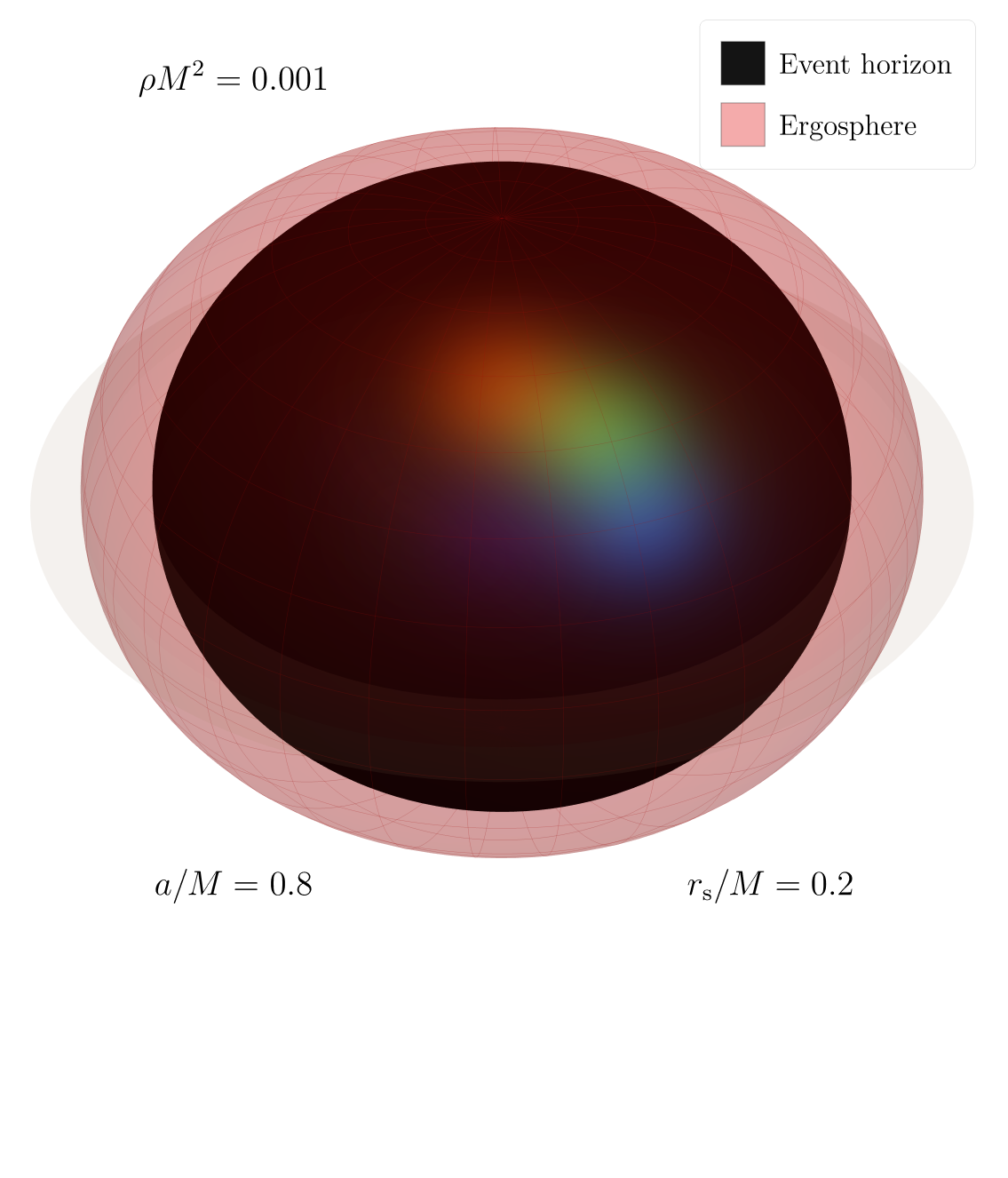}
        \includegraphics[scale=0.4]{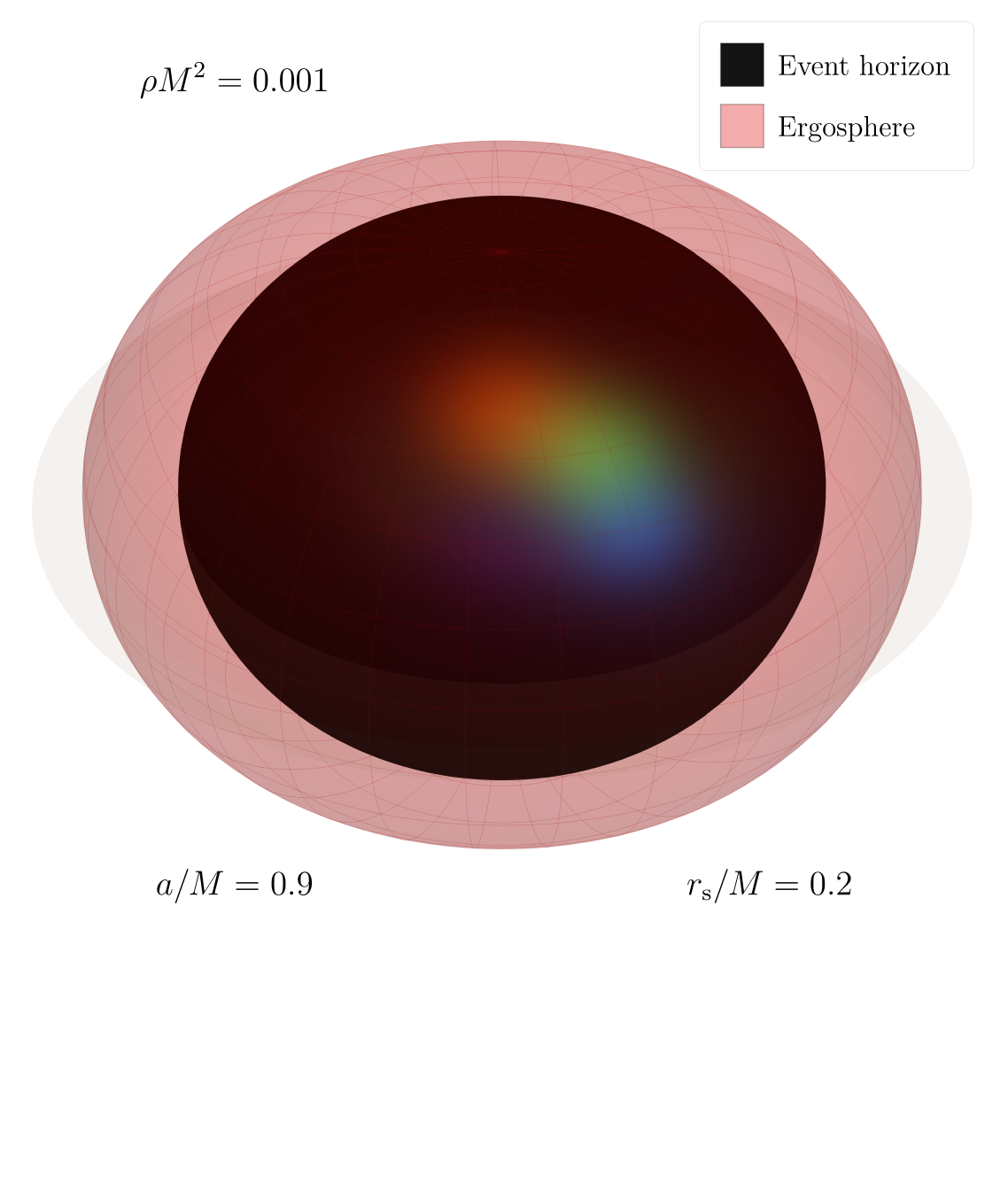}
    \caption{{Three--dimensional representations of the event horizon and ergosphere for $\rho=0.001$, a fixed halo scale radius $r_{\text{s}}/M$, and different values of the rotation parameter $a$: $a=0.6$ (top left), $a=0.7$ (top right), $a=0.8$ (bottom left), and $a=0.9$ (bottom right). The selected value of $r_{\text{s}}/M$ is stated in the panels.}}
    \label{totalones}
\end{figure}

{
In Fig.~\ref{totalones}, we plot three--dimensional representations of the event horizon and the ergosphere for $\rho=0.001$, a fixed value of $r_{\text{s}}/M$, and $a=0.6$ (top left), $a=0.7$ (top right), $a=0.8$ (bottom left), and $a=0.9$ (bottom right). }

\subsection{Frame dragging and angular velocity}

The stationary and axisymmetric character of the spacetime is associated with the Killing vector fields $\partial_{t}$ and $\partial_{\phi}$. Owing to the presence of rotation, however, an observer cannot remain static sufficiently close to the black hole. Instead, the off--diagonal component $g_{t\phi}$ forces the observer to move along the azimuthal direction, giving rise to the \textit{frame--dragging} effect.

To characterize this behavior, we consider an observer with vanishing angular momentum,
\begin{equation}
L=u_{\phi}=g_{t\phi}u^{t}+g_{\phi\phi}u^{\phi}=0.
\end{equation}
Its angular velocity with respect to an observer located at infinity is then obtained as
\begin{equation}
\omega \equiv \frac{\mathrm{d}\phi}{\mathrm{d}t} = \frac{u^{\phi}}{u^{t}} = -\frac{g_{t\phi}}{g_{\phi\phi}}.
\label{angularvelocitydefinition}
\end{equation}
For the rotating black hole surrounded by the Hernquist dark matter distribution, Eq.~\eqref{angularvelocitydefinition} gives \cite{visser2007kerr,grumiller2022black}
\begin{equation}
{ \omega(r,\theta;\rho,r_{\text{s}},a) = \frac{ a\left[r^{2}+a^{2}-\Delta(r)\right] }{ \left(r^{2}+a^{2}\right)^{2} -a^{2}\Delta(r)\sin^{2}\theta }. }
\label{omegaCompact}
\end{equation}
Notice that the factor multiplying $\Delta(r)$ in the denominator is $\sin^{2}\theta$. Substituting
\begin{equation}
{ \Delta(r) = r^{2}+a^{2}-2Mr -\frac{4\pi\rho r_{\text{s}}^{3}r^{2}}{r+r_{\text{s}}} }
\end{equation}
into Eq.~\eqref{omegaCompact}, the angular velocity can be written explicitly as
\begin{equation}
{ \omega(r,\theta;\rho,r_{\text{s}},a) = \frac{ 2aMr(r+r_{\text{s}}) +4\pi a\rho r_{\text{s}}^{3}r^{2} }{ \mathcal{D}(r,\theta;\rho,r_{\text{s}},a) }, }
\label{omegaExplicit}
\end{equation}
where
\begin{equation}
{
\begin{aligned}
\mathcal{D}(r,\theta;\rho,r_{\text{s}},a) ={}& (r+r_{\text{s}}) \left(r^{2}+a^{2}\right)^{2} \\ &- a^{2}\sin^{2}\theta \left[ (r+r_{\text{s}}) \left(r^{2}+a^{2}-2Mr\right) -4\pi\rho r_{\text{s}}^{3}r^{2} \right].
\end{aligned}}
\label{omegaDenominator}
\end{equation}

{
The Kerr result is recovered whenever the halo term is removed, either by taking $\rho\to0$ or $r_{\text{s}}\to0$. In this limit, Eq.~\eqref{omegaCompact} reduces to}
\begin{equation}
\omega_{\mathrm{Kerr}}(r,\theta) = \frac{ 2aMr }{ \left(r^{2}+a^{2}\right)^{2} -a^{2}\left(r^{2}-2Mr+a^{2}\right)\sin^{2}\theta }.
\end{equation}
Likewise, the angular velocity vanishes when $a\to0$, consistently with the absence of frame dragging in the static configuration.

At the event horizon, where $\Delta(r_{h})=0$, Eq.~\eqref{omegaCompact} assumes the simpler form
\begin{equation}
{
\Omega_{h} \equiv \omega(r_{h},\theta;\rho,r_{\text{s}},a) = \frac{a}{r_{h}^{2}+a^{2}}. }
\label{horizonAngularVelocity}
\end{equation}
{
Although Eq.~\eqref{horizonAngularVelocity} contains neither $\rho$ nor $r_{\text{s}}$ explicitly, both halo parameters enter through the corrected horizon radius $r_{h}$. The absence of angular dependence in $\Omega_{h}$ shows that the event horizon rotates with a uniform angular velocity.}

{
The separate effects of the Hernquist density and scale radius can be established directly from Eq.~\eqref{omegaCompact}. At fixed $(r,\theta,a,M,r_{\text{s}})$, we obtain }
\begin{equation}
{
\frac{\partial\omega}{\partial\rho} = \frac{ 4\pi a r_{\text{s}}^{3}r^{2} \left(r^{2}+a^{2}\right) \left(r^{2}+a^{2}\cos^{2}\theta\right) }{ (r+r_{\text{s}}) \left[ \left(r^{2}+a^{2}\right)^{2} -a^{2}\Delta(r)\sin^{2}\theta \right]^{2} }, }
\label{omegaDerivative}
\end{equation}
{
whereas, at fixed $(r,\theta,a,M,\rho)$,}
\begin{equation}
{
\frac{\partial\omega}{\partial r_{\text{s}}} = \frac{ 4\pi a\rho r_{\text{s}}^{2}r^{2} (3r+2r_{\text{s}}) \left(r^{2}+a^{2}\right) \left(r^{2}+a^{2}\cos^{2}\theta\right) }{ (r+r_{\text{s}})^{2} \left[ \left(r^{2}+a^{2}\right)^{2} -a^{2}\Delta(r)\sin^{2}\theta \right]^{2} }. }
\label{omegaRsDerivative}
\end{equation}
{
For positive $a$, $\rho$, $r_{\text{s}}$, and $r$, both derivatives are positive wherever the denominator remains regular. At a fixed exterior point, increasing either halo parameter strengthens the angular velocity measured by a zero angular momentum observer. This statement must be distinguished from the behavior of the horizon angular velocity: Eqs.~\eqref{horizonAngularVelocity} and \eqref{horizonmonotonicity} imply that, for fixed $a$, $\Omega_{h}$ decreases as either $\rho$ or $r_{\text{s}}$ increases because the outer horizon moves outward. An increase in $a$ enhances the frame--dragging effect over the range of parameters considered in the numerical analysis. }

\begin{figure}
    \centering
     \includegraphics[scale=0.48]{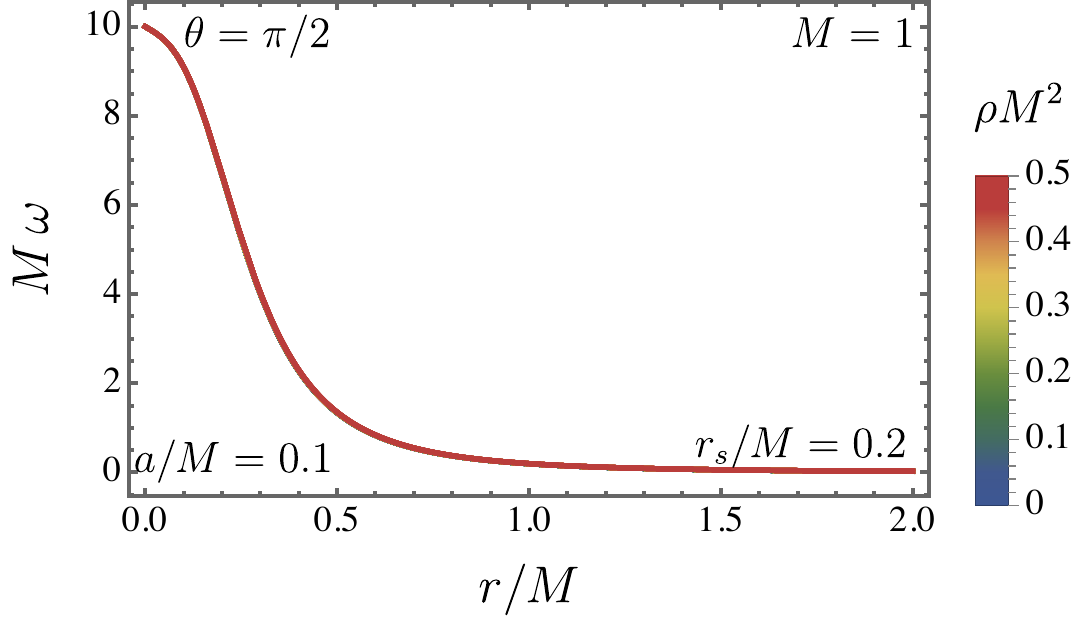}
     \includegraphics[scale=0.48]{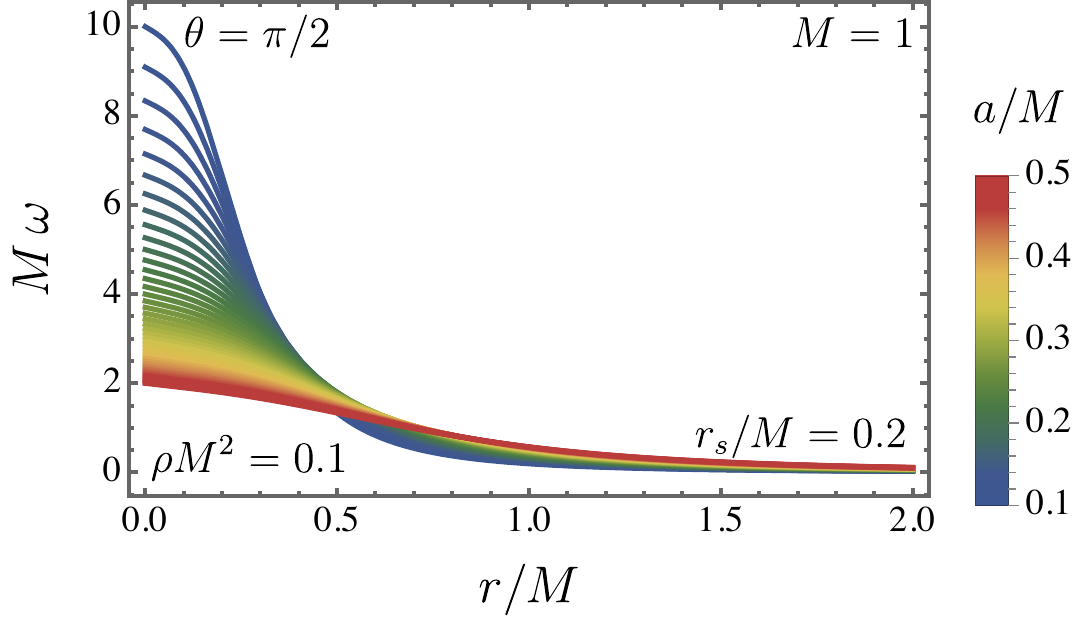}
    \caption{{Angular velocity $\omega(r,\theta;\rho,r_{\text{s}},a)$ as a function of the radial coordinate $r$ for different values of $\rho$ (left panel) and of the rotation parameter $a$ (right panel). The polar angle and the independent halo scale radius $r_{\text{s}}/M$ are fixed at the values indicated in the respective panels.}}
    \label{velangular2d}
\end{figure}

{
In Fig.~\ref{velangular2d}, we present the radial behavior of $\omega(r,\theta;\rho,r_{\text{s}},a)$ for different values of the rotation and Hernquist density parameters, together with the corresponding Kerr result ($\rho=0$). Each nonvacuum panel is a fixed $r_{\text{s}}/M$ slice of the full parameter space. The curves show that the dark matter distribution modifies the dragging of inertial frames, with larger values of $\rho$ producing higher angular velocities at fixed $r$. The dependence on $r_{\text{s}}$ is independently governed by Eq.~\eqref{omegaRsDerivative}.}

\section{Black hole thermodynamics }


\subsection{Hawking temperature and extremal configurations }

The evaluation of the surface gravity is more conveniently performed in a coordinate system that remains regular at the event horizon. For this purpose, we rewrite the rotating geometry introduced in Eq.~(\ref{newbumblebeerotating}) in terms of Eddington--Finkelstein--type coordinates \cite{christodoulou1971reversible,ruiz2019thermodynamic}. Unlike the original Boyer--Lindquist--like chart, this coordinate representation removes the apparent divergences associated with the roots of $\Delta(r)$. Accordingly, we introduce new temporal and azimuthal coordinates, denoted by $\tau$ and {$\widetilde{\varphi}$}, respectively, through the following transformations:
\ie
\mathrm{d}\tau =\mathrm{d}t+\frac{r^{2}+a^{2}}{\Delta(r)}\mathrm{d}r, \qquad { \mathrm{d}\widetilde{\varphi} =\mathrm{d}\varphi+\frac{a}{\Delta(r)}\mathrm{d}r. }
\fe
After implementing these coordinate redefinitions, the line element can be recast in the following horizon--regular form:
\ie
\begin{split}
\mathrm{d}s^{2}={}& -\frac{\Delta(r)-a^{2}\sin^{2}\theta}{\Sigma}\,\mathrm{d}\tau^{2} +2\,\mathrm{d}\tau\,\mathrm{d}r +\Sigma\,\mathrm{d}\theta^{2} -\frac{2a\left[r^{2}+a^{2}-\Delta(r)\right]\sin^{2}\theta}{\Sigma}\, \mathrm{d}\tau\,\mathrm{d}\widetilde{\varphi} \\ &-2a\sin^{2}\theta\,\mathrm{d}r\,\mathrm{d}\widetilde{\varphi} +\frac{\sin^{2}\theta}{\Sigma} \left[\left(r^{2}+a^{2}\right)^{2} -a^{2}\Delta(r)\sin^{2}\theta\right] \mathrm{d}\widetilde{\varphi}^{2}.
\end{split}
\fe
{
For any simple horizon located at $r=r_i$, the null generator is the linear combination of the stationary and axial Killing fields}
\ie
{ \chi_i =\frac{\partial}{\partial\tau} +\Omega_i\frac{\partial}{\partial\widetilde{\varphi}}, \qquad \Omega_i=\frac{a}{r_i^{2}+a^{2}}.}
\fe
{
In particular, the outer event horizon at $r=r_h$ is generated by $\chi_h$, with $\Omega_h=a/(r_h^{2}+a^{2})$. The factor multiplying the axial Killing field is therefore $a$, rather than $\Delta(r)$, which vanishes on the horizon.}

Returning to the Boyer--Lindquist coordinate system, the null generator of the outer horizon can be expressed as a linear combination of the Killing fields associated with stationarity and axial symmetry. We consider the Killing horizon $K$ on which $\chi_h^\mu$ becomes null. Since the integral curves of $\chi_h^\mu$ are null geodesics on $K$, although they are not necessarily affinely parametrized, their acceleration must remain parallel to the horizon generator. The surface gravity $\kappa_h$ is precisely the coefficient that establishes this proportionality \cite{wald2010general}, namely,
\ie
\begin{split}
\kappa_h &=\frac{\Delta'(r_h)}{2\left(r_h^{2}+a^{2}\right)} = \frac{ \left(r_h-M\right)\left(r_h+r_{\mathrm{s}}\right)^{2} -2\pi\rho\,r_h r_{\mathrm{s}}^{3}\left(r_h+2r_{\mathrm{s}}\right) }{ \left(r_h^{2}+a^{2}\right)\left(r_h+r_{\mathrm{s}}\right)^{2} }.
\end{split}
\fe
The result obtained for $\kappa_h$ provides the geometric basis for the thermodynamic analysis developed below, since it directly determines the Hawking temperature and, consequently, the remaining thermal properties of the rotating black hole.

The thermodynamic interpretation of black holes was established through the laws of black hole mechanics formulated by Bardeen, Carter, and Hawking \cite{bardeen1973four}. These relations associate the geometric properties of the event horizon with the conventional variables of thermodynamics. In this correspondence, the constancy of the surface gravity over a stationary horizon constitutes the zeroth law and plays the same role as the uniformity of temperature in a system at thermal equilibrium \cite{page2005hawking}.

Variations of the black hole parameters are governed by the first law, which relates the change in the mass to the corresponding changes in the horizon area, angular momentum, and, when present, electric charge. The terms entering this relation can be interpreted, respectively, as variations of internal energy, entropy, rotational work, and electromagnetic work \cite{carlip2014black}. {In the analysis below, the environmental parameters $\rho$ and $r_{\mathrm{s}}$ are held fixed. Allowing them to vary would require the corresponding additional work terms in an extended first law.} The second law follows from the nondecreasing character of the total horizon area in classical processes, in direct correspondence with the requirement that the entropy of an isolated thermodynamic system cannot decrease \cite{davies1978thermodynamics}. Finally, the third law states that no finite physical process can drive the surface gravity to zero, mirroring the impossibility of reaching absolute zero through a finite sequence of thermodynamic operations \cite{hawking1976black}.

The physical content of black hole thermodynamics was further clarified by the analyses of Christodoulou, who distinguished reversible from irreversible transformations in black hole dynamics \cite{christodoulou1970reversible}. In particular, these results showed that the horizon area behaves as an irreversible quantity during classical processes. This geometric interpretation acquired a thermodynamic meaning through Bekenstein's proposal that the entropy carried by a black hole must be proportional to the area of its event horizon \cite{1bekenstein2020black,2bekenstein1974generalized}. Together with the quantum description of Hawking radiation, this relation led to the \textit{Bekenstein--Hawking} entropy and established a direct connection between horizon geometry and thermodynamic entropy \cite{furtado2023thermal}. {For the perturbative formulas derived below, we assume $M^{2}>a^{2}$ and define $d\equiv\sqrt{M^{2}-a^{2}}$, and  $R\equiv M+d=r_{+}^{(0)}$. The relevant expansion parameter is dimensionless; a sufficient condition for the first--order horizon expansion to remain controlled is
\[\varepsilon_{\mathrm{h}} \equiv \frac{2\pi\rho\,r_{\mathrm{s}}^{3}R} {d\left(R+r_{\mathrm{s}}\right)} \ll1.\]
This condition also makes explicit why the expansion becomes nonuniform as $d\to0$.}

Furthermore, the Hawking temperature follows directly from the surface gravity through $T=\kappa_h/(2\pi)$. For the rotating black hole immersed in the Hernquist dark matter distribution, it is given by
\begin{equation}
\begin{aligned}
T(\rho,r_{\mathrm{s}},a,M) ={}& \frac{ \left(r_h-M\right)\left(r_h+r_{\mathrm{s}}\right)^{2} -2\pi\rho\,r_h r_{\mathrm{s}}^{3}\left(r_h+2r_{\mathrm{s}}\right) }{ 2\pi\left(r_h^{2}+a^{2}\right) \left(r_h+r_{\mathrm{s}}\right)^{2} } \\ ={}& \frac{d}{4\pi M R} +\frac{\rho\,r_{\mathrm{s}}^{3}}{2M} \left[ \frac{a^{2}}{M d\left(R+r_{\mathrm{s}}\right)} -\frac{R+2r_{\mathrm{s}}}{\left(R+r_{\mathrm{s}}\right)^{2}} \right] +\mathcal{O}\!\left(\rho^{2}\right).
\end{aligned}
\label{hawking_temperature}
\end{equation}
The first line of Eq.~\eqref{hawking_temperature} represents the complete expression, with $r_h$ determined by the full horizon equation, whereas the second line is its perturbative expansion for a weak halo contribution. This expansion assumes a nonextremal configuration, $M^{2}>a^{2}$, and should therefore not be employed arbitrarily close to the extremal regime{, as indicated explicitly by the factor $1/d$}.

The consistency of this result can be verified from its relevant limiting cases. In the absence of rotation, we obtain
\begin{equation}
\lim_{a\to0}T(\rho,r_{\mathrm{s}},a,M) =\frac{1}{8\pi M} -{ \frac{r_{\mathrm{s}}^{3}\left(M+r_{\mathrm{s}}\right)} {M\left(2M+r_{\mathrm{s}}\right)^{2}}\,\rho } +\mathcal{O}\!\left(\rho^{2}\right).
\label{temperature_static_limit}
\end{equation}
The first term corresponds to the usual Schwarzschild temperature, while the contribution proportional to $\rho$ is negative. In this manner, for a fixed black hole mass, the presence of the Hernquist dark matter distribution reduces the temperature of the static configuration.

On the other hand, removing the dark matter contribution gives
\begin{equation}
\lim_{\rho\to0}T(\rho,r_{\mathrm{s}},a,M)
=\frac{\sqrt{M^{2}-a^{2}}}
{4\pi M\left(M+\sqrt{M^{2}-a^{2}}\right)},
\label{temperature_kerr_limit}
\end{equation}
which is precisely the Hawking temperature of the Kerr black hole. The simultaneous limits $a\to0$ and $\rho\to0$ consequently lead to
\begin{equation}
\lim_{a,\rho\to0}T(\rho,r_{\mathrm{s}},a,M)
=\frac{1}{8\pi M},
\label{temperature_schwarzschild_limit}
\end{equation}
recovering the Schwarzschild result{, as required}.

At this stage, a natural question concerns the possible existence of a zero-temperature endpoint. Nevertheless, the perturbative expression presented in Eq.~\eqref{hawking_temperature} was derived under the nonextremal condition $M^{2}>a^{2}$ and becomes nonuniform when $M\to a$. In this manner, the extremal mass cannot be determined consistently by simply imposing that the perturbative expansion of the temperature vanishes. Instead, the zero-temperature configuration must be obtained from the simultaneous conditions
\begin{equation}
\Delta(r_{\mathrm{ext}})=0, \qquad \Delta'(r_{\mathrm{ext}})=0.
\label{extremal_conditions}
\end{equation}
For small values of $\rho$, these conditions yield
\begin{equation}
\begin{aligned}
r_{\mathrm{ext}} ={}& M +\frac{ 2\pi\rho\,r_{\mathrm{s}}^{3}M \left(M+2r_{\mathrm{s}}\right) }{ \left(M+r_{\mathrm{s}}\right)^{2} } +\mathcal{O}\!\left(\rho^{2}\right), \\ a_{\mathrm{ext}}^{2} ={}& M^{2} +\frac{ 4\pi\rho\,r_{\mathrm{s}}^{3}M^{2} }{ M+r_{\mathrm{s}} } +\mathcal{O}\!\left(\rho^{2}\right).
\end{aligned}
\label{extremal_quantities}
\end{equation}
Equivalently, for fixed $a$ and fixed $r_{\mathrm{s}}$, the extremal mass is
\begin{equation}
{
M_{\mathrm{ext}} =a -\frac{2\pi\rho\,a r_{\mathrm{s}}^{3}} {a+r_{\mathrm{s}}} +\mathcal{O}\!\left(\rho^{2}\right). }
\label{extremal_mass_fixed_a}
\end{equation}
The Hernquist contribution therefore shifts the zero-temperature configuration toward smaller values of the mass at fixed $a$ and $r_{\mathrm{s}}$. However, the existence of a solution satisfying $T=0$ establishes an extremal configuration and does not, by itself, prove the formation of a dynamically stable remnant. Such an interpretation would additionally require an investigation of the evaporation dynamics and the stability of the final state.

{
To display the thermodynamic quantities without assigning numerical values to dimensionful parameters, we introduce a fixed reference length $\ell_{0}$ and define
\begin{equation}
\widehat{M}=\frac{M}{\ell_{0}}, \quad \widehat{a}=\frac{a}{\ell_{0}}, \quad \widehat{r}_{\mathrm{s}}=\frac{r_{\mathrm{s}}}{\ell_{0}}, \quad \widehat{\rho}=\rho\ell_{0}^{2}, \quad \widehat{T}=\ell_{0}T, \quad \widehat{S}=\frac{S}{\ell_{0}^{2}}, \quad \widehat{C}=\frac{C}{\ell_{0}^{2}}.
\label{dimensionless_thermodynamic_variables}
\end{equation}
In this manner, $[M]=[a]=[r_{\mathrm{s}}]=L$, $[\rho]=L^{-2}$, $[T]=L^{-1}$, and the area-based quantities represented by $S$ and $C$ scale as $L^{2}$ in the conventions adopted here. The reference scale is fixed while $\widehat{M}$ is varied.
}

\begin{figure}
\centering
\includegraphics[scale=0.6]{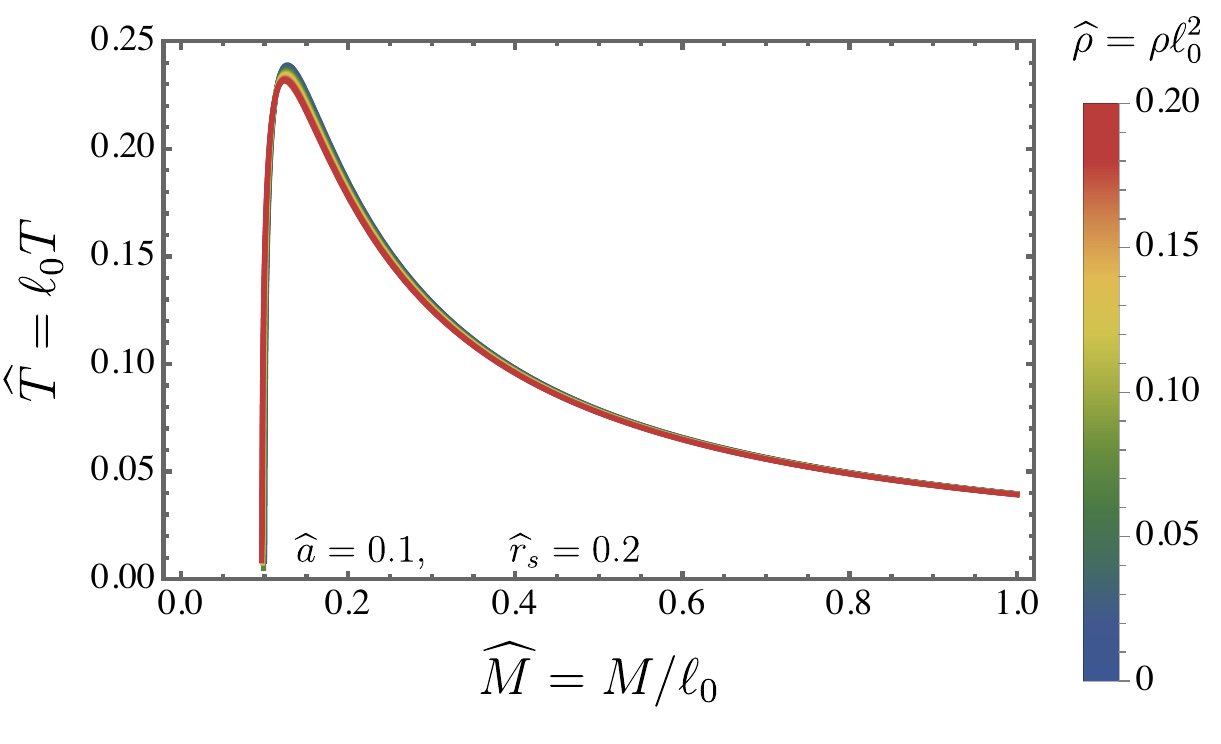}
\caption{{ Dimensionless Hawking temperature $\widehat{T}=\ell_{0}T$ as a function of $\widehat{M}=M/\ell_{0}$ for $0\leq\widehat{\rho}=\rho\ell_{0}^{2}\leq0.2$, with increments $\Delta\widehat{\rho}=0.01$. The remaining parameters are fixed at $\widehat{a}=0.1$ and $\widehat{r}_{s}=0.2$.  }}
\label{temhawww}
\end{figure}

{
The behavior of the Hawking temperature is displayed in Fig.~\ref{temhawww}. The curves were obtained from the complete, unexpanded expression for $T$, with the outer horizon determined directly from the full cubic equation and with $\widehat{r}_{\mathrm{s}}=0.2$ held fixed. Each curve begins at its own zero-temperature extremal mass. Increasing $\widehat{\rho}$ shifts this endpoint toward smaller $\widehat{M}$. At masses sufficiently above the displaced endpoints, the halo contribution suppresses the temperature; in the immediate neighborhood of the Kerr endpoint, however, the ordering reverses because the configurations with $\widehat{\rho}>0$ are already nonextremal. The curves therefore cross, and it is not correct to state that increasing $\rho$ suppresses the temperature throughout the entire displayed interval.
}

\subsection{Bekenstein--Hawking entropy}

The Bekenstein--Hawking entropy is determined by one quarter of the event horizon area. For the rotating geometry considered here, the horizon area is
\begin{equation}
A_{h}=4\pi\left(r_{h}^{2}+a^{2}\right),
\label{horizon_area}
\end{equation}
and the corresponding entropy becomes
\begin{equation}
{
S(\rho,r_{\mathrm{s}},a,M)
=\frac{A_{h}}{4}
=\pi\left(r_{h}^{2}+a^{2}\right).
}
\label{entropy_general}
\end{equation}
Using the perturbative expression for the horizon radius, the entropy can be written up to first order in $\rho$ as
\begin{equation}
{
S(\rho,r_{\mathrm{s}},a,M)
=2\pi M R
+\frac{
4\pi^{2}\rho\,r_{\mathrm{s}}^{3}R^{3}
}{
d\left(R+r_{\mathrm{s}}\right)
}
+\mathcal{O}\!\left(\rho^{2}\right),
}
\label{entropy_expanded}
\end{equation}
where $d=\sqrt{M^{2}-a^{2}}$ and $R=M+d$. The zeroth-order contribution is precisely the Kerr entropy,
\begin{equation}
\lim_{\rho\to0}S(\rho,r_{\mathrm{s}},a,M)
=2\pi M\left(M+\sqrt{M^{2}-a^{2}}\right).
\label{entropy_kerr}
\end{equation}
{The simultaneous limits $a\to0$ and $\rho\to0$ recover the Schwarzschild area law:}
\begin{equation}
\lim_{a,\rho\to0}S(\rho,r_{\mathrm{s}},a,M)
=4\pi M^{2}.
\label{entropy_schwarzschild}
\end{equation}

The correction proportional to $\rho$ in Eq.~\eqref{entropy_expanded} is positive throughout the nonextremal domain. This behavior follows directly from the outward displacement of the event horizon produced by the Hernquist distribution. Since the horizon area increases with $r_h$, the surrounding dark matter enhances the entropy and, consequently, the number of thermodynamic degrees of freedom associated with the black hole.

\begin{figure}
\centering
\includegraphics[scale=0.6]{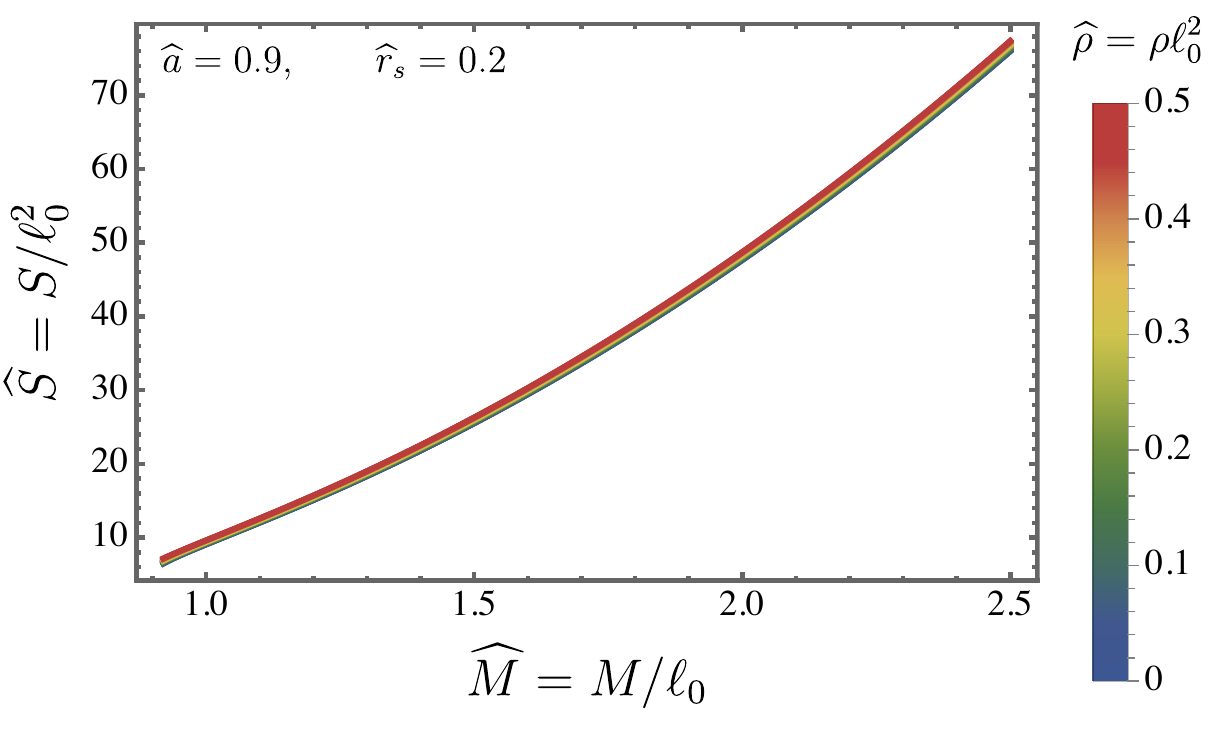}
\caption{{Dimensionless Bekenstein--Hawking entropy $\widehat{S}=S/\ell_{0}^{2}$ as a function of $\widehat{M}=M/\ell_{0}$ for $0\leq\widehat{\rho}=\rho\ell_{0}^{2}\leq0.5$, with increments $\Delta\widehat{\rho}=0.01$. The remaining parameters are fixed at $\widehat{a}=0.9$ and $\widehat{r}_{s}=0.2$.}}
\label{entroo}
\end{figure}

{
Figure~\ref{entroo} displays the dimensionless entropy as a function of $\widehat{M}$ for $\widehat{a}=0.9$, $\widehat{r}_{\mathrm{s}}=0.2$, and different values of $\widehat{\rho}$. The curves were evaluated using the complete horizon radius. At fixed mass, the entropy increases monotonically with $\rho$. For a genuinely fixed, independent $r_{\mathrm{s}}$, the large-mass expansion gives $\Delta S\simeq16\pi^{2}\rho r_{\mathrm{s}}^{3}M$, whereas $\Delta S/S_{\mathrm{Kerr}}\simeq4\pi\rho r_{\mathrm{s}}^{3}/M$. In this case, the absolute separation can grow with $M$ over the displayed range even though the fractional halo correction decreases asymptotically. This distinction would be lost if $r_{\mathrm{s}}/M$ were silently held fixed while $M$ was varied.
}

\subsection{Heat capacity and thermal stability}

{
To examine the thermal response along a definite family of solutions, we hold $a$, $\rho$, and the independent scale radius $r_{\mathrm{s}}$ fixed and define
}
\begin{equation}
{
C_{a,\rho,r_{\mathrm{s}}}(M)
=T\left(\frac{\partial S}{\partial T}\right)_{a,\rho,r_{\mathrm{s}}}
=T\,
\frac{
\left(\dfrac{\partial S}{\partial M}\right)_{a,\rho,r_{\mathrm{s}}}
}{
\left(\dfrac{\partial T}{\partial M}\right)_{a,\rho,r_{\mathrm{s}}}
}.
}
\label{heat_capacity_definition}
\end{equation}
{The notation $C_{a,\rho,r_{\mathrm{s}}}$ is preferable to $C_V$ because no thermodynamic volume is being held fixed. Moreover, fixing $a$ is not the same ensemble as fixing the angular momentum $J=aM$.} The complete analytical expression is not displayed because of its considerable length. Nevertheless, the full result, without an expansion in $\rho$, was employed throughout the numerical analysis.

In the nonrotating limit, the heat capacity can be written exactly in terms of the positive static horizon radius as
\begin{equation}
\begin{aligned}
\lim_{a\to0}C_{a,\rho,r_{\mathrm{s}}}(M) ={}& -\frac{ 2\pi r_h^{2}\left(r_h+r_{\mathrm{s}}\right) \left[ \left(r_h+r_{\mathrm{s}}\right)^{2} -4\pi\rho r_{\mathrm{s}}^{4} \right] }{ \left(r_h+r_{\mathrm{s}}\right)^{3} -4\pi\rho r_{\mathrm{s}}^{4} \left(3r_h+r_{\mathrm{s}}\right) } \\ ={}& -8\pi M^{2} -\frac{ 64\pi^{2}\rho\,r_{\mathrm{s}}^{3}M^{2} \left(4M^{2}+6Mr_{\mathrm{s}}+r_{\mathrm{s}}^{2}\right) }{ \left(2M+r_{\mathrm{s}}\right)^{3} } +\mathcal{O}\!\left(\rho^{2}\right).
\end{aligned}
\label{heat_capacity_static}
\end{equation}
{
The exact first line agrees with the general $r_{\mathrm{s}}$ static result of Ref.~\cite{Jha:2025xjf}. The negative sign of Eq.~\eqref{heat_capacity_static} shows that the nonrotating weak-halo branch remains thermally unstable, and the first-order halo correction increases the magnitude of the negative heat capacity.
}

When the dark matter contribution is removed, we obtain
\begin{equation}
\lim_{\rho\to0}C_{a,\rho,r_{\mathrm{s}}}(M)
=-\frac{
2\pi M\left(M+\sqrt{M^{2}-a^{2}}\right)^{2}
\sqrt{M^{2}-a^{2}}
}{
M\sqrt{M^{2}-a^{2}}-a^{2}
},
\label{heat_capacity_kerr}
\end{equation}
which corresponds to the Kerr heat capacity along a family with fixed rotation parameter $a$. Finally, the simultaneous limits $a\to0$ and $\rho\to0$ give
\begin{equation}
\lim_{a,\rho\to0}C_{a,\rho,r_{\mathrm{s}}}(M)
=-8\pi M^{2},
\label{heat_capacity_schwarzschild}
\end{equation}
recovering the Schwarzschild heat capacity.

\begin{figure}
\centering
\includegraphics[scale=0.6]{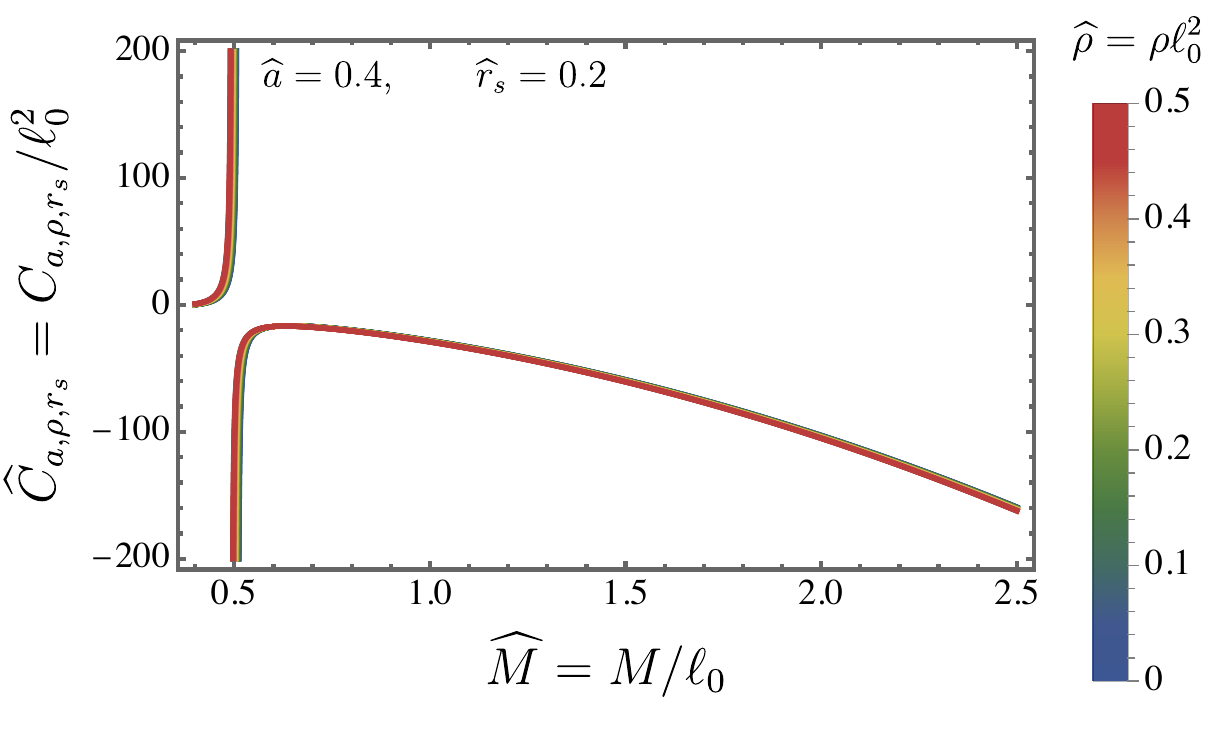}
\caption{{ Dimensionless heat capacity $\widehat{C}_{a,\rho,r_{\mathrm{s}}} =C_{a,\rho,r_{\mathrm{s}}}/\ell_{0}^{2}$ as a function of $\widehat{M}=M/\ell_{0}$ for $0\leq\widehat{\rho}=\rho\ell_{0}^{2}\leq0.5$, with increments $\Delta\widehat{\rho}=0.01$. The remaining parameters are fixed at $\widehat{a}=0.4$ and $\widehat{r}_{\mathrm{s}}=0.2$.}}
\label{heattt}
\end{figure}

{
The heat capacity is displayed in Fig.~\ref{heattt} as a function of $\widehat{M}$ for $\widehat{a}=0.4$, $\widehat{r}_{\mathrm{s}}=0.2$, and different values of $\widehat{\rho}$. Along this restricted fixed-$(a,\rho,r_{\mathrm{s}})$ family, the branch with $C_{a,\rho,r_{\mathrm{s}}}>0$ has a positive local thermal response, whereas the branch with $C_{a,\rho,r_{\mathrm{s}}}<0$ has a negative response. This statement should not be promoted to ensemble-independent stability because fixing $a$ differs from fixing $J$ or $\Omega_h$.
}

The divergence of the heat capacity occurs whenever
\begin{equation}
{
\left(
\frac{\partial T}{\partial M}
\right)_{a,\rho,r_{\mathrm{s}}}
=0,
}
\label{critical_condition}
\end{equation}
and therefore corresponds to an extremum of the Hawking temperature. Across this point, the heat capacity changes sign and the local thermal response of the selected family is modified. {For the parameters of Fig.~\ref{heattt}, increasing $\widehat{\rho}$ shifts the pole toward smaller $\widehat{M}$. The divergence can be described as a Davies--type response function singularity. Establishing a continuous phase transition, however, would require a thermodynamically consistent ensemble and the corresponding potential; the pole alone is not sufficient.}

\section{Quantum tunneling}
\label{sec:quantum_tunnelling}

We now investigate the quantum emission of massless particles from the rotating black hole immersed in the Hernquist dark matter distribution. Contrary to the Kerr case, the horizon function of the present geometry is not quadratic in the radial coordinate. In this way, the corresponding tunnelling calculation cannot be expressed in terms of the standard Kerr quantities $r_{+}$ and $r_{-}$. Instead, the pole associated with the event horizon must be determined directly from the radial function $\Delta(r)$ and its derivative evaluated at the outer horizon.

{The line element obtained in Sec.~II is written in Boyer--Lindquist coordinates in Eq.~\eqref{newbumblebeerotating}.} Unlike the Kerr geometry, the function $\Delta(r)$ cannot generally be written as a product involving only two physical roots. This difference directly modifies the residue entering the tunnelling probability.

To remove the coordinate singularity at $\Delta(r_h)=0$, we introduce the advanced coordinates
\begin{equation}
\mathrm{d}v = \mathrm{d}t+\frac{r^{2}+a^{2}}{\Delta(r)}\,\mathrm{d}r, \qquad \mathrm{d}\bar{\varphi} = \mathrm{d}\varphi+\frac{a}{\Delta(r)}\,\mathrm{d}r.
\label{advanced_transformation_tunnelling}
\end{equation}
The metric then assumes the form
\begin{equation}
\begin{split}
\mathrm{d}s^{2}={}& -\frac{\Delta(r)-a^{2}\sin^{2}\theta}{\Sigma}\,\mathrm{d}v^{2} +2\,\mathrm{d}v\,\mathrm{d}r +\Sigma\,\mathrm{d}\theta^{2} -2a\sin^{2}\theta\,\mathrm{d}r\,\mathrm{d}\varphi \\ &-\frac{2a\left[r^{2}+a^{2}-\Delta(r)\right]\sin^{2}\theta}{\Sigma} \,\mathrm{d}v\,\mathrm{d}\varphi +\frac{\left(r^{2}+a^{2}\right)^{2} -a^{2}\Delta(r)\sin^{2}\theta}{\Sigma} \sin^{2}\theta\,\mathrm{d}\varphi^{2}.
\end{split}
\label{metric_tunnelling_advanced}
\end{equation}
This coordinate system is regular at the outer event horizon and allows the outgoing and ingoing solutions to be distinguished without introducing an additional co-rotating coordinate transformation.

The inverse metric components required in the Hamilton--Jacobi equation are
\begin{equation}
\begin{aligned}
g^{vv} &= \frac{a^{2}\sin^{2}\theta}{\Sigma}, & g^{vr} &= \frac{r^{2}+a^{2}}{\Sigma}, & g^{v\varphi} &= \frac{a}{\Sigma}, \\ g^{rr} &= \frac{\Delta(r)}{\Sigma}, & g^{r\varphi} &= \frac{a}{\Sigma}, & g^{\theta\theta} &= \frac{1}{\Sigma}, & g^{\varphi\varphi} &= \frac{1}{\Sigma\sin^{2}\theta}.
\end{aligned}
\label{inverse_metric_tunnelling}
\end{equation}

Within the semiclassical approximation, the leading contribution to the tunnelling process is governed by the classical action $I$, which satisfies the Hamilton--Jacobi equation for a massless particle,
\begin{equation}
g^{\mu\nu} \partial_{\mu}I \partial_{\nu}I =0.
\label{HJ_tunnelling}
\end{equation}
Since the geometry is independent of $v$ and $\varphi$, we consider the ansatz
$I = -\omega v +J\varphi +\mathcal{F}(r,\theta)$, where $\omega$ and $J$ denote the conserved energy and azimuthal angular momentum of the emitted particle, respectively. Notice that $J$ is not identified with the angular momentum of the black hole and remains an independent conserved quantity.

Substituting Eq.~\eqref{inverse_metric_tunnelling} into Eq.~\eqref{HJ_tunnelling}, we obtain
\begin{equation}
\begin{split}
0={}& \Delta(r)\mathcal{F}_{r}^{\,2} + 2\left[ aJ-\left(r^{2}+a^{2}\right)\omega \right]\mathcal{F}_{r} + \left( J\csc\theta -a\omega\sin\theta \right)^{2} + \mathcal{F}_{\theta}^{\,2},
\end{split}
\label{HJ_expanded_tunnelling}
\end{equation}
where
$\mathcal{F}_{r} = \frac{\partial\mathcal{F}}{\partial r}$, and  $\mathcal{F}_{\theta} = \frac{\partial\mathcal{F}}{\partial\theta}$.
Solving Eq.~\eqref{HJ_expanded_tunnelling} for the radial derivative gives
\begin{equation}
\mathcal{F}_{r}^{(\pm)} = \frac{ \mathcal{K}(r) \pm \sqrt{ \mathcal{K}^{2}(r) -\Delta(r)\mathcal{A}(r,\theta) } } {\Delta(r)},
\label{radial_action_tunnelling}
\end{equation}
where $\mathcal{K}(r) = \left(r^{2}+a^{2}\right)\omega-aJ$ and $\mathcal{A}(r,\theta) = \left( J\csc\theta -a\omega\sin\theta \right)^{2} + \mathcal{F}_{\theta}^{\,2}$.

Equation~\eqref{radial_action_tunnelling} contains an explicit dependence on the polar coordinate. Nevertheless, the angular contribution is multiplied by $\Delta(r)$ inside the square root. Since the tunnelling process is governed by the near-horizon regime, where $\Delta(r_h)=0$, this term does not contribute to the residue responsible for the imaginary part of the action. Consequently, it is not necessary to impose a particular value of $\theta$.

For a nonextremal configuration, the expansion of the horizon function around $r=r_h$ is
\begin{equation}
\Delta(r) = \Delta'(r_h)(r-r_h) + \mathcal{O}\left[(r-r_h)^{2}\right].
\label{delta_near_horizon_tunnelling}
\end{equation}
Using this expansion in Eq.~\eqref{radial_action_tunnelling}, the outgoing solution behaves as
\begin{equation}
\mathcal{F}_{r}^{\mathrm{out}} \simeq \frac{ 2\left[ \left(r_h^{2}+a^{2}\right)\omega-aJ \right] } { \Delta'(r_h)(r-r_h) },
\label{outgoing_solution_tunnelling}
\end{equation}
whereas the numerator of the ingoing branch vanishes at the horizon. Therefore, the ingoing solution remains regular, while the outgoing solution develops a simple pole at $r=r_h$.

The radial contribution to the outgoing action becomes
\begin{equation}
\mathcal{F}_{\mathrm{out}} \simeq \frac{ 2\left[ \left(r_h^{2}+a^{2}\right)\omega-aJ \right] } {\Delta'(r_h)} \int \frac{\mathrm{d}r}{r-r_h}.
\label{outgoing_integral_tunnelling}
\end{equation}
Regularizing the pole according to the Feynman prescription gives
\begin{equation}
\operatorname{Im}I_{\mathrm{out}} = \frac{ 2\pi\left[ \left(r_h^{2}+a^{2}\right)\omega-aJ \right] } {\Delta'(r_h)}.
\label{imaginary_action_tunnelling}
\end{equation}

The Killing vector that generates the outer event horizon is $\chi = \partial_{v} + \Omega_h\partial_{\varphi}$, where the angular velocity of the horizon is $\Omega_h = \frac{a}{r_h^{2}+a^{2}}$. {Although this expression contains neither $\rho$ nor $r_s$ explicitly, both Hernquist parameters remain encoded in the modified horizon radius $r_h$.}

In this manner, the imaginary part of the action can be expressed as
\begin{equation}
\operatorname{Im}I_{\mathrm{out}} = \frac{ 2\pi\left(r_h^{2}+a^{2}\right) } {\Delta'(r_h)} \left( \omega-\Omega_hJ \right).
\label{imaginary_action_omega_tunnelling}
\end{equation}
The combination $\omega-\Omega_hJ$ appears directly from the Hamilton--Jacobi equation. Therefore, the use of an additional co-rotating coordinate transformation is not required.

The tunnelling rate is determined by
$\Gamma = \exp\left[ -2\operatorname{Im}I_{\mathrm{out}} \right]$, which yields
\begin{equation}
\Gamma = \exp\left[ -\frac{ 4\pi\left(r_h^{2}+a^{2}\right) } {\Delta'(r_h)} \left( \omega-\Omega_hJ \right) \right].
\label{gamma_tunnelling}
\end{equation}
By comparison with the thermal distribution
$\Gamma = \exp\left[ -\beta_H \left( \omega-\Omega_hJ \right) \right]$,
we identify
\begin{equation}
\beta_H = \frac{ 4\pi\left(r_h^{2}+a^{2}\right) } {\Delta'(r_h)}.
\label{beta_tunnelling}
\end{equation}
The Hawking temperature obtained from the tunnelling process is consequently
$T = \frac{ \Delta'(r_h) } { 4\pi\left(r_h^{2}+a^{2}\right) }$.

{For the general horizon function
$\Delta(r)=r^{2}+a^{2}-2Mr-4\pi\rho r_s^{3}r^{2}/(r+r_s)$, its derivative is}
\begin{equation}
{
\Delta'(r) = 2(r-M) - \frac{4\pi\rho r_s^{3}r(r+2r_s)}{(r+r_s)^{2}}.
}
\label{delta_derivative_tunnelling}
\end{equation}
Evaluating this expression at $r=r_h$, the imaginary part of the action becomes
\begin{equation}
{
\operatorname{Im}I_{\mathrm{out}}
=
\frac{\pi\left(r_h^{2}+a^{2}\right)\left(\omega-\Omega_hJ\right)}
{(r_h-M)-\dfrac{2\pi\rho r_s^{3}r_h(r_h+2r_s)}{(r_h+r_s)^{2}}}.
}
\label{imaginary_action_explicit_tunnelling}
\end{equation}
Accordingly, the tunnelling rate associated with the new geometry is
\begin{equation}
{
\Gamma
=
\exp\left[
-\frac{2\pi\left(r_h^{2}+a^{2}\right)\left(\omega-\Omega_hJ\right)}
{(r_h-M)-\dfrac{2\pi\rho r_s^{3}r_h(r_h+2r_s)}{(r_h+r_s)^{2}}}
\right].
}
\label{gamma_explicit_tunnelling}
\end{equation}

In possession of the tunnelling rate in Eq.~\eqref{gamma_explicit_tunnelling}, we can estimate the mean number of particles created in each mode. For this purpose, it is convenient to introduce
\begin{equation}
{
\mathcal{D}_h
=
(r_h-M)
-\frac{2\pi\rho r_s^{3}r_h(r_h+2r_s)}{(r_h+r_s)^{2}}
=\frac{\Delta'(r_h)}{2}.
}
\label{Dh_particle_creation}
\end{equation}
The tunnelling probability can then be written as
\begin{equation}
\Gamma = \exp\left[ -\frac{ 2\pi(r_h^{2}+a^{2}) } {\mathcal{D}_h} \left(\omega-\Omega_hJ\right) \right].
\label{gamma_compact_particle_creation}
\end{equation}
By recalling that the Hawking temperature obtained from the tunnelling process is
\begin{equation}
T = \frac{ \mathcal{D}_h } { 2\pi(r_h^{2}+a^{2}) },
\label{temperature_particle_creation}
\end{equation}
Eq.~\eqref{gamma_compact_particle_creation} assumes the thermal form
\begin{equation}
\Gamma = \exp\left[ -\frac{ \omega-\Omega_hJ } {T} \right].
\label{gamma_thermal_particle_creation}
\end{equation}

\begin{figure}
    \centering
    \includegraphics[scale=0.55]{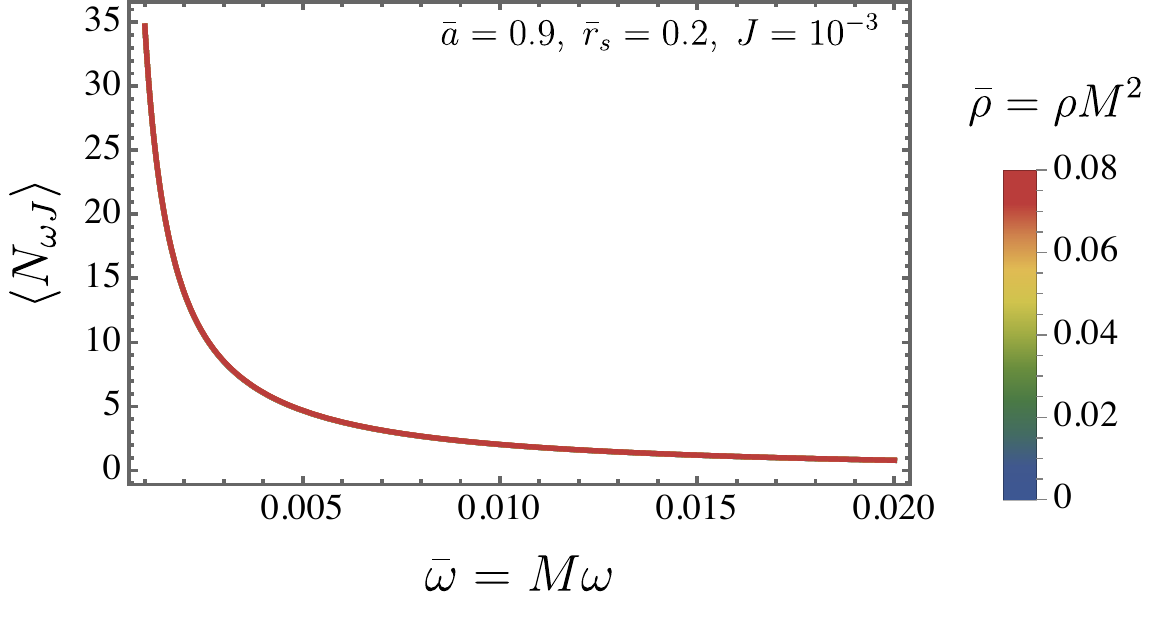}
    \includegraphics[scale=0.55]{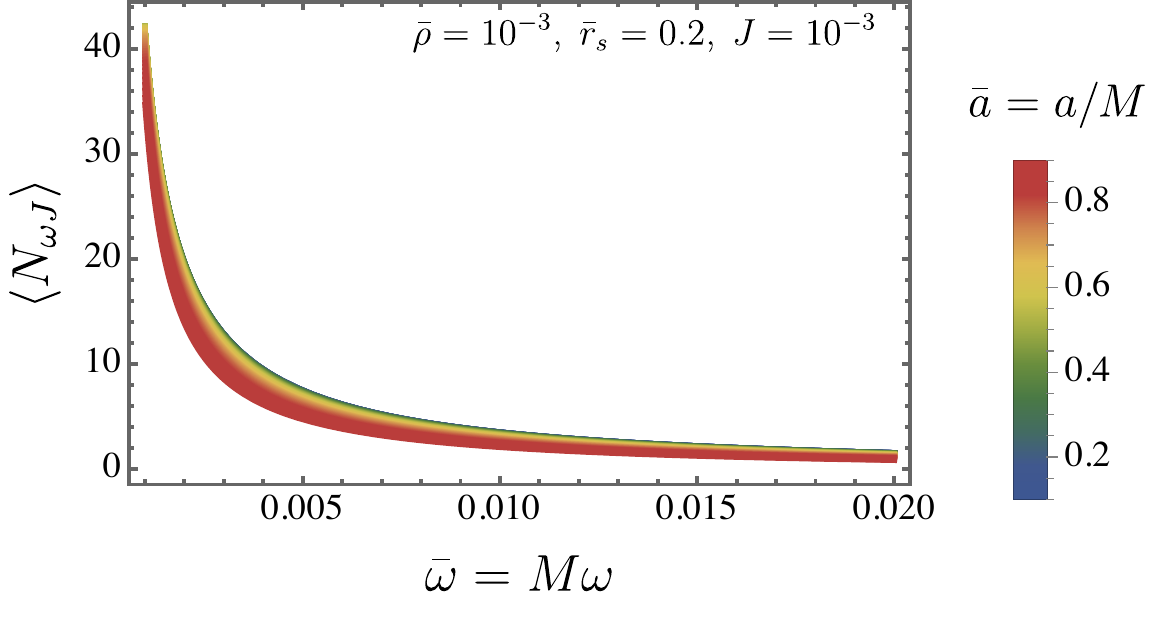}
    \caption{{Particle occupation number $\left\langle N_{\omega J}\right\rangle$ as a function of the dimensionless frequency $\bar{\omega}=M\omega$, calculated with the complete outer horizon for $\bar{r}_s=0.2$ and $J=10^{-3}$. The top panel varies $\bar{\rho}=\rho M^{2}$ at fixed $\bar{a}=0.9$, whereas the bottom panel varies $\bar{a}=a/M$ at fixed $\bar{\rho}=10^{-3}$.}}
    \label{particheflefmd}
\end{figure}

The quantity $\Gamma$ represents the relative probability associated with the emission of a single particle. To determine the mean occupation number, we consider the probability of creating $n$ identical bosonic particles in a mode characterized by $(\omega,J)$. The corresponding probability distribution can be written as
\begin{equation}
P_n = \left(1-\Gamma\right)\Gamma^{n}, \qquad \sum_{n=0}^{\infty}P_n=1,
\label{Pn_particle_creation}
\end{equation}
provided that $\omega-\Omega_hJ>0$. The mean number of particles occupying this mode is consequently
\begin{equation}
\left\langle N_{\omega J}\right\rangle = \sum_{n=0}^{\infty}nP_n = \frac{\Gamma}{1-\Gamma}.
\label{mean_number_gamma}
\end{equation}
Using Eq.~\eqref{gamma_thermal_particle_creation}, we obtain
\begin{equation}
\left\langle N_{\omega J}\right\rangle = \frac{1}{ \exp\left[ \dfrac{ 2\pi(r_h^{2}+a^{2}) }{ \mathcal{D}_h } \left(\omega-\Omega_hJ\right) \right] -1}
\label{mean_number_particle_creation}
\end{equation}
Equivalently, the occupation number assumes the familiar Bose--Einstein form
\begin{equation}
\left\langle N_{\omega J}\right\rangle = \frac{1}{ \exp\left[ \dfrac{\omega-\Omega_hJ}{T} \right] -1 }.
\label{Bose_Einstein_particle_creation}
\end{equation}
Therefore, the angular velocity of the event horizon acts as a chemical potential for the azimuthal angular momentum of the emitted particle. {The dark matter contribution, controlled independently by $\rho$ and $r_s$, affects the particle spectrum through both the corrected horizon radius $r_h$ and the modified temperature $T$. To display the spectrum using dimensionless quantities, we define $\bar{\omega}=M\omega$, $\bar{\rho}=\rho M^{2}$, $\bar{a}=a/M$, and $\bar{r}_s=r_s/M$. In Fig.~\ref{particheflefmd}, we fix $\bar{r}_s=0.2$ and evaluate $r_h$ from the complete cubic horizon equation. For the high--spin choice $\bar{a}=0.9$ in the upper panel, increasing $\bar{\rho}$ produces a mild enhancement of the occupation number over the displayed frequency range. This behavior is consistent with the nonuniform ordering of the Hawking temperature curves discussed in Sec.~IV. In the lower panel, the high--spin curves are suppressed over most of the displayed range as $\bar{a}$ increases at fixed $\bar{\rho}=10^{-3}$; very close to the low--frequency edge, however, the competition between $T$ and the rotational chemical potential $\Omega_hJ$ can make the ordering nonmonotonic. }

The occupation number in Eq.~\eqref{Bose_Einstein_particle_creation} gives the number of particles created in an individual mode. To obtain a spectral particle density, it must be combined with the density of states. In the blackbody approximation, for $g$ independent polarization or internal degrees of freedom, the number density per unit frequency is
\begin{equation}
\frac{\mathrm{d}n}{\mathrm{d}\omega} = \frac{ g\omega^{2} } {2\pi^{2}} \frac{1}{ \exp\left[ \dfrac{\omega-\Omega_hJ}{T} \right] -1 }.
\label{spectral_number_density}
\end{equation}
Substituting the explicit temperature, this expression becomes
\begin{equation}
{
\frac{\mathrm{d}n}{\mathrm{d}\omega}
=
\frac{g\omega^{2}}{2\pi^{2}}
\left\{
\exp\left[
\frac{2\pi(r_h^{2}+a^{2})}
{(r_h-M)-\dfrac{2\pi\rho r_s^{3}r_h(r_h+2r_s)}{(r_h+r_s)^{2}}}
\left(\omega-\Omega_hJ\right)
\right]-1
\right\}^{-1}.
}
\label{spectral_number_density_explicit}
\end{equation}

\begin{figure}
    \centering
    \includegraphics[scale=0.6]{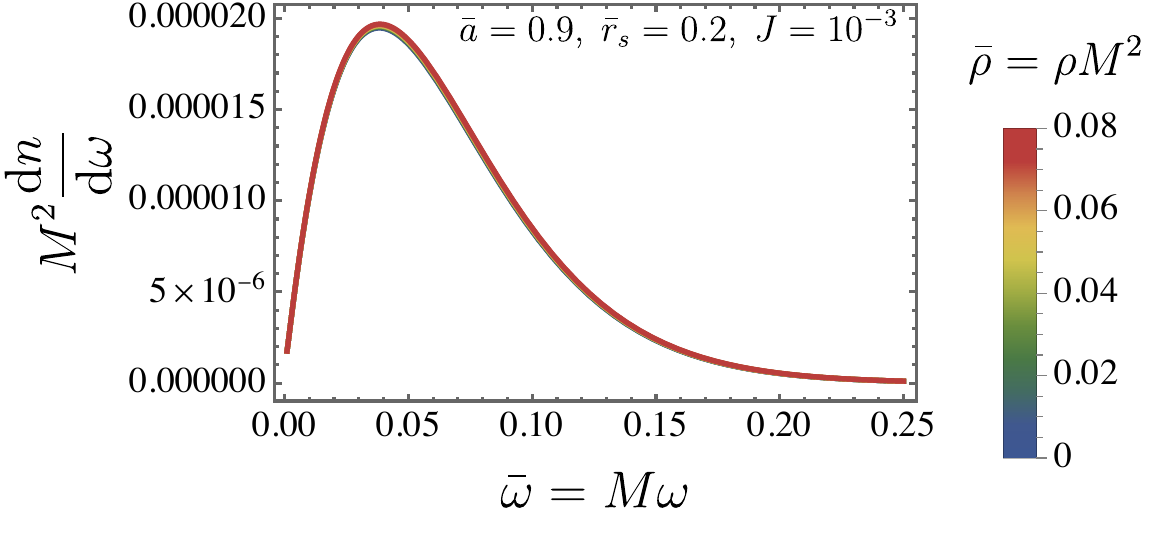}
    \includegraphics[scale=0.6]{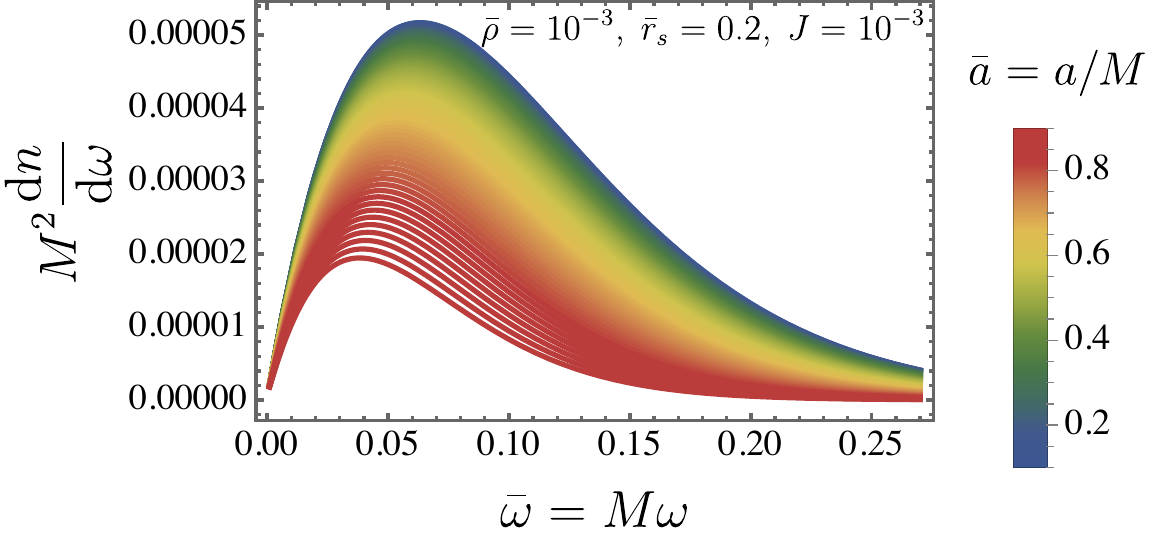}
    \caption{{Dimensionless particle creation density $M^{2}\mathrm{d}n/\mathrm{d}\omega$ as a function of $\bar{\omega}=M\omega$, calculated with the complete outer horizon for $\bar{r}_s=0.2$, $J=10^{-3}$, and $g=1$. The top panel varies $\bar{\rho}$ at fixed $\bar{a}=0.9$, whereas the bottom panel varies $\bar{a}$ at fixed $\bar{\rho}=10^{-3}$.}}
    \label{dndw}
\end{figure}

Equation~\eqref{spectral_number_density_explicit} should be understood as a thermal estimate. {In Fig.~\ref{dndw}, we plot the dimensionless spectral density $M^{2}\mathrm{d}n/\mathrm{d}\omega$ as a function of $\bar{\omega}=M\omega$. Owing to the phase--space factor proportional to $\omega^{2}$, the spectrum vanishes in the very low--frequency limit, reaches a maximum, and then decreases at higher frequencies. For the high--spin parameters of the upper panel, increasing $\bar{\rho}$ mildly raises the spectral density. In the lower panel, increasing $\bar{a}$ lowers the spectrum over most of the displayed interval, with a possible nonmonotonic ordering near the low--frequency edge owing to the rotational chemical potential. }

The actual number of particles reaching infinity is modified by the potential barrier outside the event horizon. Introducing the greybody factor $\mathcal{T}_{\ell m}(\omega)$, the physically relevant number flux is
\begin{equation}
\frac{\mathrm{d}^{2}N}{\mathrm{d}t\,\mathrm{d}\omega} = \frac{1}{2\pi} \sum_{\ell,m} \frac{ \mathcal{T}_{\ell m}(\omega) } { \exp\left[ \dfrac{\omega-m\Omega_h}{T} \right] -1 },
\label{number_flux_greybody}
\end{equation}
where we have identified the azimuthal angular momentum of each mode with $J=m$ in natural units. Subsequently, the energy flux is
\begin{equation}
\frac{\mathrm{d}^{2}E}{\mathrm{d}t\,\mathrm{d}\omega} = \frac{\omega}{2\pi} \sum_{\ell,m} \frac{ \mathcal{T}_{\ell m}(\omega) } { \exp\left[ \dfrac{\omega-m\Omega_h}{T} \right] -1 }.
\label{energy_flux_greybody}
\end{equation}
Therefore, the tunnelling method determines the thermal factor, while the greybody factor controls the fraction of particles that can propagate through the exterior effective potential.

\begin{figure}
    \centering
    \includegraphics[scale=0.6]{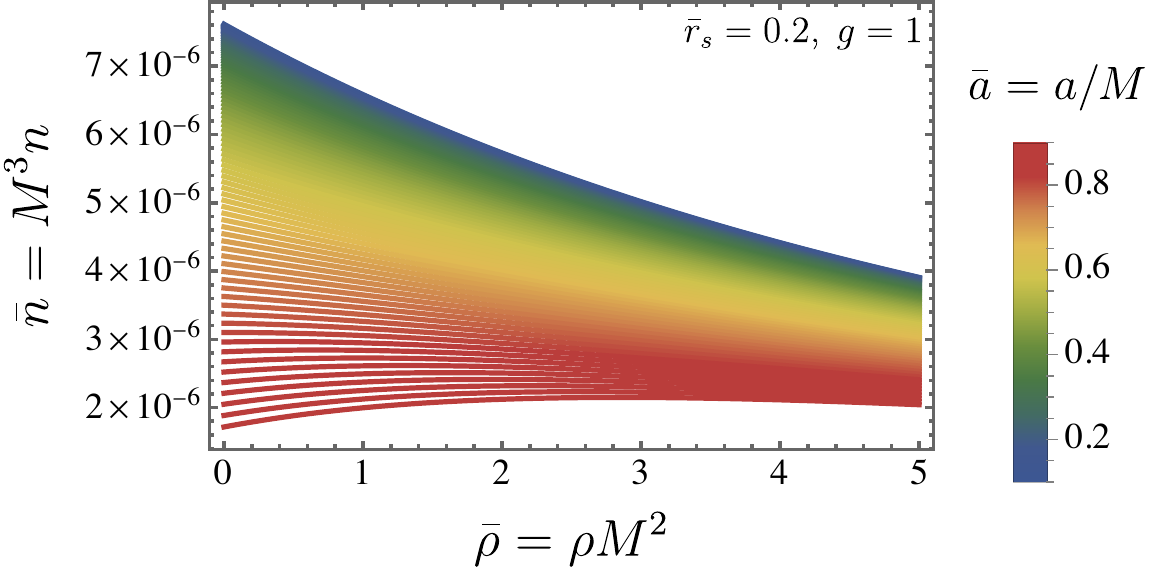}
    \caption{{Dimensionless particle number density $\bar{n}=M^{3}n$ as a function of $\bar{\rho}=\rho M^{2}$ for different values of $\bar{a}=a/M$, calculated from the complete outer horizon with $\bar{r}_s=0.2$ and $g=1$.}}
    \label{nparticle}
\end{figure}

A simpler estimate of the total particle number density can be obtained by considering modes with vanishing azimuthal angular momentum, $J=0$, and neglecting the greybody corrections. In this approximation,
\begin{equation}
n = \frac{g}{2\pi^{2}} \int_{0}^{\infty} \frac{ \omega^{2}\,\mathrm{d}\omega } { \exp\left(\omega/T\right)-1 }.
\label{total_number_density_integral}
\end{equation}
Using
\begin{equation}
\int_{0}^{\infty} \frac{x^{2}\,\mathrm{d}x}{e^{x}-1} = 2\zeta(3),
\end{equation}
we find
\begin{equation}
n = \frac{ g\zeta(3) } {\pi^{2}} T^{3}.
\label{total_number_density_temperature}
\end{equation}
For the black hole under consideration, this gives
\begin{equation}
{
n
=
\frac{g\zeta(3)}{8\pi^{5}}
\frac{
\left[
(r_h-M)
-\dfrac{2\pi\rho r_s^{3}r_h(r_h+2r_s)}{(r_h+r_s)^{2}}
\right]^{3}
}
{(r_h^{2}+a^{2})^{3}}.
}
\label{particle_creation_density_explicit}
\end{equation}
This result shows that, within the blackbody approximation, the particle creation density scales with the third power of the Hawking temperature. In this manner, even a relatively small correction to $T$ can produce a more pronounced modification in the number of particles created. {In Fig.~\ref{nparticle}, we display the dimensionless density $\bar{n}=M^{3}n$ as a function of $\bar{\rho}$ for different values of $\bar{a}$, keeping $\bar{r}_s=0.2$ and $g=1$. Since $\bar{n}\propto (MT)^{3}$, the curves reproduce the nonuniform temperature ordering discussed in Sec.~IV: increasing $\bar{\rho}$ suppresses the density for the lower-- and moderate--spin branches, whereas the ordering may reverse for the high--spin branches. }

For instance, in the nonrotating and small $\rho$ regime, the Hawking temperature is
\begin{equation}
{
T
=
\frac{1}{8\pi M}
-\frac{\rho r_s^{3}(M+r_s)}
{M(2M+r_s)^{2}}
+\mathcal{O}(\rho^{2}).
}
\nonumber
\end{equation}
The corresponding particle-creation density becomes
\begin{equation}
{
n
=
\frac{g\zeta(3)}{512\pi^{5}M^{3}}
\left[
1
-\frac{24\pi\rho r_s^{3}(M+r_s)}
{(2M+r_s)^{2}}
+\mathcal{O}(\rho^{2})
\right].
}
\label{static_particle_density}
\end{equation}
{Thus, for positive $\rho$ and $r_s$, the Hernquist contribution suppresses the particle density in the static limit at fixed $M$ and independent $r_s$.} This behavior is a direct consequence of the reduction in the Hawking temperature caused by the surrounding dark matter distribution.

For the rotating configuration, the spectrum also depends on the effective energy
$\widetilde{\omega} = \omega-\Omega_hJ$.
For modes satisfying $\widetilde{\omega}>0$, the occupation number is finite and describes ordinary Hawking emission. As $\omega$ approaches $\Omega_hJ$ from above, the mean occupation number increases. However, when $\omega<\Omega_hJ$, the simple interpretation of $\Gamma$ as an ordinary probability is no longer valid. This interval corresponds to the bosonic superradiant regime. In the complete emission spectrum, the absorption probability $\mathcal{T}_{\ell m}(\omega)$ also changes sign in this region, ensuring that the physical flux in Eq.~\eqref{number_flux_greybody} remains well defined.

It is also important to emphasize that Eq.~\eqref{particle_creation_density_explicit} represents an ideal thermal estimate rather than the exact number of particles observed at infinity. A complete determination of the particle production rate requires the calculation of the greybody factors for the present geometry, followed by the summation over the angular modes in Eq.~\eqref{number_flux_greybody}. Nevertheless, the tunnelling result already establishes how {the independent Hernquist parameters $\rho$ and $r_s$, together with} the black hole rotation modify the thermal population of the emitted particles.


\section{Hawking emission and evaporation dynamics}
\label{sec:evaporation}

{The tunnelling calculation fixes the thermal factor of the emitted modes, whereas the flux at infinity also depends on their transmission through the exterior potential. We therefore separate the analytical Stefan--Boltzmann estimate from the mode--resolved treatment presented below.}

Before calculating such transmission coefficients, an analytical estimate of the evaporation rate can be obtained by treating the black hole as an imperfect thermal emitter. In natural units, the Stefan--Boltzmann approximation gives
\begin{equation}
\mathcal{L}_{\mathrm{SB}} \equiv -\frac{\mathrm{d}M}{\mathrm{d}t} = \frac{\pi^{2}}{120}\, g_{\star}\, \varepsilon_{\mathrm{em}}\, A_{h}{T^{4}},
\label{SB_luminosity}
\end{equation}
where $g_{\star}$ denotes the effective number of radiated degrees of freedom and $\varepsilon_{\mathrm{em}}$ is a frequency--averaged emissivity. The ideal blackbody limit is recovered for $\varepsilon_{\mathrm{em}}=1$. In the present approximation, the factor $\varepsilon_{\mathrm{em}}$ phenomenologically accounts for the difference between the horizon area and the frequency--dependent absorption cross section. {Once the greybody factors are known, Eq.~\eqref{SB_luminosity} must be replaced by the mode-resolved energy flux in Sec.~\ref{subsec:spectral_emission_rate}.}


\subsection{Expansion for a weak Hernquist contribution}
\label{subsec:weak_rho_evaporation}

We initially preserve the complete dependence on the rotation parameter and expand the luminosity only in the Hernquist parameter, {holding the independent scale radius $r_s$ fixed}. For convenience, we introduce
\begin{equation}
x(M,a) = \sqrt{1-\frac{a^{2}}{M^{2}}}, \qquad 0<x\leq 1.
\label{x_evaporation}
\end{equation}
Using the horizon radius, horizon area, and Hawking temperature previously obtained, Eq.~\eqref{SB_luminosity} becomes
\begin{equation}
{
\mathcal{L}_{\mathrm{SB}} = \frac{ g_{\star}\varepsilon_{\mathrm{em}} }{ 3840\pi M^{2} } \frac{x^{4}}{(1+x)^{3}} \left[ 1+\rho\,\mathcal{C}_{\rho}(M,a,r_s) \right] + \mathcal{O}(\rho^{2}),}
\label{luminosity_small_rho}
\end{equation}
where
\begin{equation}
{
\mathcal{C}_{\rho}(M,a,r_s) =\frac{2\pi r_s^{3}R}{d(R+r_s)^{2}} \left[ \frac{R(R+r_s)}{M} +\frac{4a^{2}(R+r_s)}{Md} -4(R+2r_s) \right],}
\label{rho_luminosity_correction}
\end{equation}
The zeroth--order contribution in Eq.~\eqref{luminosity_small_rho} corresponds to the Stefan--Boltzmann luminosity of the Kerr black hole when the horizon area is adopted as the effective emitting area.

For fixed $a$ and $\rho$, {with $r_s$ also held fixed}, the time required for the mass to decrease from $M_i$ to $M_f$ is consequently given by
\begin{equation}
{
\begin{aligned}
t_{i\rightarrow f} ={}& \frac{3840\pi}{ g_{\star}\varepsilon_{\mathrm{em}} } \int_{M_f}^{M_i} M^{2} \frac{(1+x)^{3}}{x^{4}} \left[ 1-\rho\,\mathcal{C}_{\rho}(M,a,r_s) \right] \mathrm{d}M + \mathcal{O}(\rho^{2}),
\end{aligned}}
\label{lifetime_small_rho}
\end{equation}
where $x=x(M,a)$. Equation~\eqref{lifetime_small_rho} provides the evaporation timescale to first order in $\rho$ while maintaining the complete rotational dependence. For selected values of $M_i$, $M_f$, $a$, and $\rho$, this integral can be evaluated numerically.

Although an analytical primitive can be obtained after introducing a suitable rationalizing variable, the resulting expression involves several rational and logarithmic contributions and does not provide a particularly transparent physical interpretation. A more compact analytical result can be obtained by additionally considering the slowly rotating regime.

{The correction $\mathcal{C}_{\rho}(M,a,r_s)$ need not have a fixed sign throughout the subextremal parameter space. It is negative in the slowly rotating regime, so that the Hernquist contribution suppresses the luminosity there, but it may change sign at larger spin. Its divergence as $x\rightarrow 0$ also makes the small $\rho$ expansion nonuniform near extremality; hence Eq.~\eqref{luminosity_small_rho} is restricted to the perturbative domain established by the horizon expansion.}


\subsection{Weak halo and slow rotation regime}
\label{subsec:slow_rotation_evaporation}

{We now impose $a^{2}/M^{2}\ll 1$ together with the weak-halo condition at fixed $r_s$. Expanding Eq.~\eqref{luminosity_small_rho} through the first nonvanishing rotational order gives}
\begin{equation}
{
\mathcal{L}_{\mathrm{SB}} =\frac{\gamma}{M^{2}} \left[ 1-\frac{5a^{2}}{4M^{2}} -\rho\,\mathcal{A}(M,r_s) +\rho a^{2}\mathcal{B}(M,r_s) \right] +\mathcal{O}\left(\rho^{2},\frac{a^{4}}{M^{4}}, \frac{\rho a^{4}}{M^{2}}\right), }
\label{slow_rotation_luminosity}
\end{equation}
where $\gamma \equiv \frac{ g_{\star}\varepsilon_{\mathrm{em}} }{ 30720\pi }$, {and}
\begin{equation}
{
\mathcal{A}(M,r_s) =\frac{8\pi r_s^{3}(2M+3r_s)}{(2M+r_s)^{2}}, \qquad \mathcal{B}(M,r_s) =\frac{2\pi r_s^{3} \left(40M^{2}+50Mr_s+19r_s^{2}\right)} {M^{2}(2M+r_s)^{3}}. }
\label{slow_rotation_coefficients}
\end{equation}
{Both $\mathcal{A}$ and $\mathcal{B}$ are positive. Thereby, the independent $a^{2}$ and $\rho$ corrections suppress the luminosity, while the mixed term partially compensates for that suppression within the perturbative regime.}

To determine the lifetime, we invert the mass loss equation. Retaining terms up to the same perturbative order, we obtain
\begin{equation}
{
\frac{\mathrm{d}t}{\mathrm{d}M} = -\frac{1}{\gamma} \left[ M^{2}+\frac{5a^{2}}{4} +\frac{8\pi\rho M^{2}r_s^{3}(2M+3r_s)} {(2M+r_s)^{2}} +\frac{2\pi\rho a^{2}r_s^{4}(30M+11r_s)} {(2M+r_s)^{3}} \right]. }
\label{inverse_mass_loss}
\end{equation}
The coefficient of the mixed contribution in Eq.~\eqref{inverse_mass_loss} differs from the corresponding coefficient in Eq.~\eqref{slow_rotation_luminosity} because the inversion of the luminosity also generates a product between the independent rotational and Hernquist corrections.

{For a compact presentation of the integrated result, we define}
\begin{equation}
{
\begin{aligned}
\mathcal{F}_{0}(M;r_s) ={}&\frac{\pi r_s^{3}}{2}(2M+r_s)^{2} -3\pi r_s^{5}\ln\!\left(\frac{2M+r_s}{r_s}\right) -\frac{2\pi r_s^{6}}{2M+r_s},\\ \mathcal{F}_{2}(M;r_s) ={}&\pi r_s^{4} \left[ -\frac{15}{2M+r_s} +\frac{2r_s}{(2M+r_s)^{2}} \right]. \end{aligned} }
\label{fixed_a_primitives}
\end{equation}
{Integration between an initial mass $M_i$ and a final mass $M_f$ then gives}
\begin{equation}
{
\begin{aligned}
t_{i\rightarrow f}^{(a=\mathrm{const.})} ={}&\frac{1}{\gamma} \Bigg\{ \frac{M_i^{3}-M_f^{3}}{3} +\frac{5a^{2}}{4}(M_i-M_f) +\rho\!\left[\mathcal{F}_{0}(M_i;r_s) -\mathcal{F}_{0}(M_f;r_s)\right]\\ &\quad +\rho a^{2}\!\left[\mathcal{F}_{2}(M_i;r_s) -\mathcal{F}_{2}(M_f;r_s)\right] \Bigg\} +\mathcal{O}\left(a^{4},\rho^{2},\rho a^{4}\right).
\end{aligned}}
\label{fixed_a_lifetime}
\end{equation}
Equation~\eqref{fixed_a_lifetime} constitutes a closed analytical estimate of the evaporation time containing the leading rotational, Hernquist, and mixed corrections.

Nevertheless, this result must be interpreted as a partial evaporation timescale. If $a$ is maintained fixed while the mass decreases, the ratio $a/M$ progressively increases. The slow--rotation approximation therefore eventually ceases to be valid. The final mass employed in Eq.~\eqref{fixed_a_lifetime} must consequently satisfy $\frac{a^{2}}{M_f^{2}}\ll 1$. For any nonvanishing fixed value of $a$, extending Eq.~\eqref{fixed_a_lifetime} to $M_f=0$ would violate the approximation used in its derivation.

In the nonrotating limit, Eq.~\eqref{fixed_a_lifetime} reduces to
\begin{equation}
{
t_{i\rightarrow f}^{(a=0)} =\frac{1}{\gamma} \left[ \frac{M_i^{3}-M_f^{3}}{3} +\rho\!\left(\mathcal{F}_{0}(M_i;r_s) -\mathcal{F}_{0}(M_f;r_s)\right) \right] +\mathcal{O}(\rho^{2}). }
\label{static_lifetime}
\end{equation}
In this particular case, the formal continuation to $M_f=0$ gives
\begin{equation}
{
t_{\mathrm{evap}}^{(a=0)} =\frac{1}{\gamma} \left[ \frac{M_i^{3}}{3} +\rho\!\left( \mathcal{F}_{0}(M_i;r_s) +\frac{3\pi r_s^{5}}{2} \right) \right] +\mathcal{O}(\rho^{2}).}
\label{static_total_lifetime}
\end{equation}
{Because $\mathcal{F}_{0}'(M;r_s)=M^{2}\mathcal{A}(M,r_s)>0$, a positive Hernquist parameter increases the evaporation timescale in the weak--halo regime.}


\subsection{Lifetime along a constant spin trajectory}
\label{subsec:constant_spin_lifetime}

A simple estimate of the complete lifetime for a slowly rotating configuration can be obtained by introducing the dimensionless rotation parameter $\chi \equiv \frac{a}{M}$, and $\chi^{2}\ll 1,$ and assuming that $\chi$ remains constant during the evaporation process. Under this assumption, $a(t)=\chi M(t)$, and the slow-rotation condition is preserved while the mass decreases.

Substituting $a=\chi M$ into Eq.~\eqref{slow_rotation_luminosity}, we find
\begin{equation}
{
\mathcal{L}_{\mathrm{SB}} =\frac{\gamma}{M^{2}} \left[ 1-\frac{5}{4}\chi^{2} -\rho\,\mathcal{A}(M,r_s) +\rho\chi^{2}M^{2}\mathcal{B}(M,r_s) \right] +\mathcal{O}\left(\chi^{4},\rho^{2},\rho\chi^{4}\right).}
\label{constant_spin_luminosity}
\end{equation}
After inversion, the corresponding evolution equation becomes
\begin{equation}
{
\frac{\mathrm{d}t}{\mathrm{d}M} = - \frac{M^{2}}{\gamma} \left[ 1+\frac{5}{4}\chi^{2} +\rho\,\mathcal{A}(M,r_s) +\rho\chi^{2}\mathcal{E}(M,r_s) \right], \qquad \mathcal{E}(M,r_s) =\frac{2\pi r_s^{4}(30M+11r_s)} {(2M+r_s)^{3}}.}
\label{constant_spin_inverse}
\end{equation}
{Introduce the remaining primitive}
\begin{equation}
{
\mathcal{F}_{\chi}(M;r_s) =\frac{\pi r_s^{4}}{4} \left[ 15(2M+r_s) -34r_s\ln\!\left(\frac{2M+r_s}{r_s}\right) -\frac{23r_s^{2}}{2M+r_s} +\frac{2r_s^{3}}{(2M+r_s)^{2}} \right],}
\label{constant_spin_primitive}
\end{equation}
{for which $\mathcal{F}_{\chi}'(M;r_s)=M^{2}\mathcal{E}(M,r_s)$. The partial lifetime is therefore}
\begin{equation}
{
\begin{aligned}
t_{i\rightarrow f}^{(\chi=\mathrm{const.})} ={}&\frac{1}{\gamma} \Bigg\{ \left(1+\frac{5}{4}\chi^{2}\right) \frac{M_i^{3}-M_f^{3}}{3} +\rho\!\left[\mathcal{F}_{0}(M_i;r_s) -\mathcal{F}_{0}(M_f;r_s)\right]\\ &\quad +\rho\chi^{2}\!\left[\mathcal{F}_{\chi}(M_i;r_s) -\mathcal{F}_{\chi}(M_f;r_s)\right] \Bigg\} +\mathcal{O}\left(\chi^{4},\rho^{2},\rho\chi^{4}\right).
\end{aligned}}
\label{constant_spin_partial_lifetime}
\end{equation}
{The formal continuation to $M_f=0$ gives}
\begin{equation}
{
\begin{aligned}
t_{\mathrm{evap}}^{(\chi=\mathrm{const.})} ={}&\frac{1}{\gamma} \Bigg\{ \left(1+\frac{5}{4}\chi^{2}\right)\frac{M_i^{3}}{3} +\rho\!\left[ \mathcal{F}_{0}(M_i;r_s) +\frac{3\pi r_s^{5}}{2} \right]\\ &\quad +\rho\chi^{2}\!\left[ \mathcal{F}_{\chi}(M_i;r_s) +\frac{3\pi r_s^{5}}{2} \right] \Bigg\}+\mathcal{O}\left(\chi^{4},\rho^{2},\rho\chi^{4}\right).
\end{aligned}}
\label{constant_spin_lifetime}
\end{equation}
{For photon emission in the ideal blackbody benchmark, one simply sets $g_{\star}=2$ and $\varepsilon_{\mathrm{em}}=1$ in these expressions.}

The assumption of a constant $\chi$ must be understood as an analytical benchmark rather than as a complete description of the evaporation of a rotating black hole. Hawking radiation carries both energy and angular momentum, and the actual trajectory in the $(M,a)$ parameter space must be determined by solving the mass and angular momentum balance equations simultaneously. In particular, bosonic superradiant modes may lead to an initial spin--down stage that is absent from the Stefan--Boltzmann approximation.

{Throughout the quasistatic sequence, $\rho$ and the independent scale radius $r_s$ are held fixed as external halo parameters. The formal integration to $M_f=0$ is controlled only if the weak-halo expansion remains valid along the entire trajectory; its limiting form requires $\rho r_s^{2}\ll 1$, with $\chi$ also sufficiently far from extremality.}

Finally, the suppression of the luminosity or the approach to a zero temperature extremal boundary does not, by itself, demonstrate the formation of a dynamically stable remnant. Such a conclusion would require the complete greybody spectrum, the coupled evolution of the mass and angular momentum, and an independent analysis of the stability of the possible endpoint.


\subsection{\texorpdfstring{{Spectral Hawking fluxes and balance equations}}{Spectral Hawking fluxes and balance equations}}
\label{subsec:spectral_emission_rate}

{For a particle species $i$ with spin $s_i$ and internal degeneracy $g_i$, the mode resolved particle emission rate is}
\begin{equation}
\frac{\mathrm{d}^{2}N_i} {\mathrm{d}t\,\mathrm{d}\omega} = \frac{g_i}{2\pi} \sum_{\ell,m} \frac{ \Gamma_{\ell m}^{(i)}(\omega) }{ \exp\left[ \dfrac{\omega-m\Omega_h}{T} \right] - (-1)^{2s_i} },
\label{particle_emission_rate}
\end{equation}
where $\Gamma_{\ell m}^{(i)}(\omega)$ denotes the greybody factor associated with the mode $(\ell,m)$, while $\Omega_h$ is the angular velocity of the event horizon. The denominator reduces to the Bose--Einstein distribution for integer spin particles and to the Fermi--Dirac distribution for half integer-spin particles.

The corresponding spectral energy emission rate is
\begin{equation}
\frac{\mathrm{d}^{2}E_i} {\mathrm{d}t\,\mathrm{d}\omega} = \frac{g_i\omega}{2\pi} \sum_{\ell,m} \frac{ \Gamma_{\ell m}^{(i)}(\omega) }{ \exp\left[ \dfrac{\omega-m\Omega_h}{T} \right] - (-1)^{2s_i} }.
\label{energy_emission_rate}
\end{equation}
Similarly, the angular momentum flux carried by the emitted particles is given by
\begin{equation}
\frac{\mathrm{d}^{2}J_i} {\mathrm{d}t\,\mathrm{d}\omega} = \frac{g_i}{2\pi} \sum_{\ell,m} \frac{ m\,\Gamma_{\ell m}^{(i)}(\omega) }{ \exp\left[ \dfrac{\omega-m\Omega_h}{T} \right] - (-1)^{2s_i} }.
\label{angular_momentum_emission_rate}
\end{equation}
Consequently, the mass and angular momentum of the black hole satisfy the balance relations
\begin{equation}
-\frac{\mathrm{d}M}{\mathrm{d}t} = \sum_i \int_{0}^{\infty} \frac{\mathrm{d}^{2}E_i} {\mathrm{d}t\,\mathrm{d}\omega} \,\mathrm{d}\omega, \qquad -\frac{\mathrm{d}{J_{\mathrm{BH}}}}{\mathrm{d}t} = \sum_i \int_{0}^{\infty} \frac{\mathrm{d}^{2}J_i} {\mathrm{d}t\,\mathrm{d}\omega} \,\mathrm{d}\omega.
\label{mass_angular_momentum_balance}
\end{equation}
These equations show that the evolution of a rotating black hole is, in general, governed by a coupled system for $M(t)$ and {$J_{\mathrm{BH}}(t)=a(t)M(t)$}. In particular, the spectrum depends on the effective frequency $\omega-m\Omega_h$, rather than only on $\omega$. For bosonic modes satisfying $\omega<m\Omega_h$, the emission enters the superradiant regime. In this case, the signed absorption probability must be employed consistently in Eq.~\eqref{particle_emission_rate}.

Before obtaining the individual greybody factors, an approximate spectrum can be introduced by treating the horizon as an imperfect bosonic blackbody. Within the same horizon area approximation {used in Eq.~\eqref{SB_luminosity}}, the spectral particle and energy fluxes become
\begin{equation}
\frac{\mathrm{d}^{2}N_{\mathrm{SB}}} {\mathrm{d}t\,\mathrm{d}\omega} = \frac{ g_{\star}\varepsilon_{\mathrm{em}}A_h }{ 8\pi^{2} } \frac{ \omega^{2} }{ \exp\left(\omega/T\right)-1 },
\label{SB_particle_spectrum}
\end{equation}
and
\begin{equation}
\frac{\mathrm{d}^{2}E_{\mathrm{SB}}} {\mathrm{d}t\,\mathrm{d}\omega} = \frac{ g_{\star}\varepsilon_{\mathrm{em}}A_h }{ 8\pi^{2} } \frac{ \omega^{3} }{ \exp\left(\omega/T\right)-1 },
\label{SB_energy_spectrum}
\end{equation}
respectively. The total particle emission rate obtained from Eq.~\eqref{SB_particle_spectrum} is
\begin{equation}
\dot{N}_{\mathrm{SB}} = \frac{ g_{\star}\varepsilon_{\mathrm{em}}\zeta(3) }{ 4\pi^{2} } A_hT^{3},
\label{total_particle_emission_rate}
\end{equation}
where $\zeta(3)$ is the Apéry constant. Integration of Eq.~\eqref{SB_energy_spectrum} reproduces the Stefan--Boltzmann luminosity given in Eq.~\eqref{SB_luminosity}. Therefore, the mean energy carried by each emitted quantum in this approximation is
\begin{equation}
\left\langle\omega\right\rangle_{\mathrm{SB}} = \frac{\mathcal{L}_{\mathrm{SB}}} {\dot{N}_{\mathrm{SB}}} = \frac{\pi^{4}}{30\zeta(3)}T \approx 2.701\,T.
\label{mean_emitted_energy}
\end{equation}

Equations~\eqref{SB_particle_spectrum}--\eqref{mean_emitted_energy} should be understood as analytical benchmarks. They neglect the {mode-dependent} transmission coefficients, the rotational chemical potential $m\Omega_h$, and the superradiant contribution. Once the effective potentials and greybody factors are explicitly determined, the complete emission spectrum must instead be evaluated from Eqs.~\eqref{particle_emission_rate} and \eqref{energy_emission_rate}.


\section{Conclusion}\label{Sec:Conclusion}

In this work, we investigated a rotating black hole immersed in a Hernquist dark matter halo. Starting from the static and spherically symmetric solution, we constructed the corresponding rotating geometry by using the noncomplexification formulation of the Newman--Janis algorithm. The resulting spacetime was characterized by the rotation parameter $a$ and by the {independent Hernquist halo parameters $\rho$ and $r_s$}, and it consistently recovered the static Hernquist configuration in the limit $a\to 0$ and the Kerr geometry when the surrounding matter contribution was removed.

We first analyzed the horizon structure of the rotating solution. In contrast with the Kerr case, the horizon equation was governed by a cubic polynomial once the Hernquist contribution was included. Since the exact roots did not lead to compact expressions, we developed a perturbative expansion for small $\rho$. {This analysis showed that, for positive $\rho$ and $r_s$, the outer event horizon was displaced toward larger radial coordinates, with its radius increasing monotonically with either halo parameter in the nonextremal domain.} The extremal configuration was also modified by the dark matter distribution, {and, at fixed $a$ and $r_s$, the zero--temperature boundary was shifted toward smaller masses}. The stationary limit surfaces were obtained from the condition $g_{tt}=0$, and the corresponding ergoregion was shown to depend on {the three independent parameters $a$, $\rho$, and $r_s$}. As expected, the ergoregion vanished in the nonrotating limit, whereas rotation produced the characteristic separation between the event horizon and the outer stationary limit surface.

We also examined the rotational properties of the spacetime through the angular velocity of zero angular momentum observers. The off diagonal metric component $g_{t\phi}$ led to the frame dragging effect, and the resulting angular velocity reduced to the usual Kerr expression when {the halo contribution was removed, either through $\rho\to0$ or $r_s\to0$}. {At fixed exterior points, increasing either $\rho$ or $r_s$ raised the angular velocity of zero angular momentum observers. By contrast, at fixed $a$, the horizon angular velocity decreased as either halo parameter increased because the outer horizon was displaced outward.} At the event horizon, the angular velocity became independent of the polar coordinate, indicating that the horizon rotated uniformly.

The thermodynamic sector was then developed from the surface gravity evaluated at the outer horizon. From this quantity, we derived the Hawking temperature and verified that the Kerr and Schwarzschild limits were recovered when the appropriate parameters were removed. {The Hernquist contribution lowered the Hawking temperature in the static weak--halo limit. For rotating configurations, however, the ordering of the temperature curves was nonuniform: the temperature was suppressed away from the shifted extremal endpoints, whereas near--extremal and high--spin scenarios could display a reversed ordering.} The Bekenstein--Hawking entropy was obtained from the horizon area, and the {halo correction proportional to $\rho$, at fixed independent $r_s$,} was positive in the nonextremal regime. Therefore, the surrounding matter distribution enlarged the event horizon and increased the entropy. The heat capacity was also analyzed {along fixed $(a,\rho,r_s)$ parameters, and its divergences separated branches with positive and negative local thermal responses. These divergences were identified as Davies--type response--function singularities; they did not, by themselves, establish continuous phase transitions or ensemble--independent thermodynamic stability.}

Furthermore, we studied the quantum emission process through the Hamilton--Jacobi tunneling method. The radial part of the action acquired an imaginary contribution from the pole at the event horizon, and the tunneling probability was written in terms of the effective energy $\omega-\Omega_h J$. This result led to the corresponding occupation number and to a {thermal} estimate of the particle creation density. {Both $\rho$ and $r_s$ entered the particle spectrum through the modified horizon radius and Hawking temperature.} In the nonrotating and weak halo regime, the Hernquist contribution suppressed the particle density, which followed directly from the reduction of the Hawking temperature. {For rotating configurations, the emission spectrum depended on the horizon angular velocity and inherited the parameter--dependent ordering of the Hawking temperature, whereas the regime $\omega<m\Omega_h$ corresponded to the bosonic superradiant sector.}

Finally, we estimated the Hawking luminosity and the evaporation time by treating the black hole as an imperfect thermal emitter within a Stefan--Boltzmann approximation. {In the weak--Hernquist and slow--rotation regime at fixed independent $r_s$, both the rotational correction and the halo correction proportional to $\rho$ reduced the luminosity, while the mixed rotational--halo contribution partially compensated for this suppression.} The corresponding {partial} evaporation time increased in the presence of a weak Hernquist halo, showing that the surrounding matter distribution delayed the mass loss process within the approximation employed. {The fixed $a$ and constant spin lifetime formulas were therefore interpreted as controlled analytical benchmarks. A complete evolution requires the simultaneous mass and angular momentum balance, together with the greybody factors and superradiant contributions. Moreover, neither the suppression of the luminosity nor the existence of a shifted zero temperature boundary was sufficient to establish the formation of a dynamically stable remnant.} The spectral particle and energy emission rates were also {formulated analytically, whereas the complete spectrum requires the mode dependent greybody factors and transmission coefficients}.

As a further perspective, it would be interesting to investigate an axisymmetric dark matter halo configuration for the magnetically charged black hole reported in Ref. \cite{Jha:2025cqf}, constructed through the {noncomplexification formulation of the} Newman--Janis technique in a manner similar to the procedure adopted in the present work. In addition, the analysis of the corresponding optical properties {of the spacetime in Ref. \cite{Jha:2025cqf}}, as well as the black hole studied here, constitutes a natural continuation of the present investigation. In particular, the study of geodesic motion, gravitational lensing, black hole shadows, and related observational signatures may provide further constraints on the role played by the dark matter halo. {These questions constitute natural directions for future investigations.}


\section*{Acknowledgments}
\hspace{0.5cm} A.A.A.F. is supported by Conselho Nacional de Desenvolvimento Cient\'{\i}fico e Tecnol\'{o}gico (CNPq) and Fundação de Apoio à Pesquisa do Estado da Paraíba (FAPESQ), project numbers 150223/2025-0 and 1951/2025. N. H is supported by the Conselho Nacional de Desenvolvimento Científico e Tecnológico (CNPq), grant No. 152891/2025-0. N. H. also acknowledges the networking support provided by COST Action CA22113 -- Fundamental challenges in theoretical physics (Theory and Challenges), CA21106 -- COSMIC WISPers in the Dark Universe: Theory, astrophysics and experiments (CosmicWISPers), CA21136 -- Addressing observational tensions in cosmology with systematics and fundamental physics (CosmoVerse), and CA23130 -- Bridging high and low energies in search of quantum gravity (BridgeQG).

\section*{Data Availability Statement}

Data Availability Statement: No Data associated with the manuscript

\bibliographystyle{ieeetr}
\bibliography{main}

\end{document}